\documentclass[journal=jctcce,manuscript=article,layout=traditional]{achemso}

\usepackage{amsmath,amssymb}
\usepackage{graphicx}
\usepackage{placeins}
\usepackage{booktabs}
\usepackage{enumitem}
\usepackage{xcolor}
\usepackage{setspace}
\usepackage{hyperref}
\SectionNumbersOn

\newcommand{\om}{\omega}
\newcommand{\G}{\mathbf{G}}
\newcommand{\A}{\mathbf{A}}
\newcommand{\jj}{\mathbf{J}}
\newcommand{\rr}{\mathbf{r}}
\newcommand{\R}{\mathbf{R}}
\newcommand{\kk}{\mathbf{k}}
\newcommand{\kgam}{\mathbf{k}_\gamma}
\newcommand{\dEabs}{\Delta E_{\mathrm{abs}}}

\newcommand{\todo}[1]{\textcolor{red}{\textbf{TODO: #1}}}

\newcommand{\LasVegas}{Department of Physics \&{} Astronomy, University of Nevada Las Vegas, Las Vegas, Nevada 89154, USA}

\newcommand{\Argonne}{X-ray Sciences Division, Argonne National Laboratory, Lemont, Illinois, 60439, USA}

\newcommand{\UCSD}{Department of Nanoengineering, University of California San Diego, La Jolla, CA 92037, USA}

\author{Daniel~Schacher}
\affiliation{\LasVegas}

\author{Tod~A.~Pascal}
 \affiliation{\UCSD}

\author{Craig~P.~Schwartz}
\affiliation{\LasVegas}
\altaffiliation{\Argonne}
\email{craig.schwartz@unlv.edu}

\author{Keith~V.~Lawler}
\affiliation{\LasVegas}
\email{keith.lawler@unlv.edu}

\title[Beyond-Dipole RT-TDDFT in FHI-aims]
{Full Minimal Coupling All-Electron Real-Time TDDFT for X-Ray-Matter Interactions}


\begin{document}

\begin{abstract}
Standard real-time time-dependent density functional theory (RT-TDDFT) couples light to matter through a spatially uniform (dipole) vector potential, an approximation that breaks down in the X-ray regime, where the photon wavelength approaches interatomic scales.
We present an all-electron, numeric-atom-centered-orbital implementation of full minimal coupling in FHI-aims that retains the spatial structure of the vector potential in both the paramagnetic and diamagnetic couplings, implements a reciprocal-space-resolved current diagnostic, and extends the propagation to periodic systems through a perturbative photon-momentum sideband coupling that recovers the density response at the photon wavevector.
Four benchmarks ranging in energy from $100$\,eV to $5$\,keV validate the implementation and exercise its reach.
The retained photon momentum proves physically consequential throughout, from measurable beyond-dipole corrections to the enabling of channels forbidden in the dipole limit.
On an all-electron footing, the implementation lays the groundwork for a general \textit{ab initio} treatment of X-ray light--matter interaction across molecules, surfaces, and solids.
\end{abstract}

\begin{tocentry}
\todo{If you have any ideas for this lmk.}
\end{tocentry}

\section{Introduction}

The development of high-harmonic-generation (HHG) sources and free-electron lasers (FELs) has transformed X-ray and extreme-ultraviolet (XUV) science;
their combination of high brightness, coherence, and ultrashort pulse duration has enabled experiments in both the linear regime (most prominently pump-probe spectroscopy with the X-ray being the probe) and, increasingly when using FELs, the nonlinear regime, including techniques such as X-ray two-photon absorption~\cite{tamasaku2014,doumy2011}, sequential multiphoton inner-shell ionization~\cite{young2010}, inner-shell X-ray lasing~\cite{rohringer2012}, and X-ray--optical wave mixing~\cite{glover2012}, and stimulated X-ray Raman scattering~\cite{tanaka2002,weninger2013}, that would otherwise be inaccessible at lower field strength. 
Interpreting both linear and nonlinear X-ray measurements has motivated the development and adaptation of a range of \textit{ab initio} methods for the XUV/X-ray range.
Among the most formally complete for two bodies is the Bethe--Salpeter equation (BSE)~\cite{onida2002}, which delivers accurate core-excitation spectra by treating the electron--core-hole interaction explicitly~\cite{shirley1998,vinson2011,vinson2012,vinson2022}, typically built on a preceding density-functional-theory (DFT) calculation when used to study X-ray spectroscopy. Spectroscopic properties can also be obtained from time-dependent density functional theory (TDDFT), in both its linear-response~\cite{casida1995} and real-time~\cite{yabana1996,li2020} formulations; TDDFT is widely used because it offers a favorable balance of accuracy and cost~\cite{marques2004,lopata2012,fernando2015,pemmaraju2019,kadek2015,lam2018,berger2021,uzundal2021,schwartz2021,helk2021,hoffmann2022,schacher2026}.
Real-time TDDFT (RT-TDDFT) is fundamentally distinct from the linear-response variant: a single nonperturbative time propagation captures the electronic response to arbitrary order when in the appropriate gauge~\cite{lopata2011}, giving simultaneous access to the linear and nonlinear response at high photon energies~\cite{yabana1996,castro2004,yabana2012,takimoto2007}, a decisive advantage for the intense, coherent fields delivered by modern light sources~\cite{tancognedejean2017}, albeit at the cost of being more expensive computationally.

Standard RT-TDDFT implementations couple the field to matter through a spatially uniform (dipole) vector potential, $\A(\rr,t)\,\rightarrow\,\A(t)$~\cite{yabana1996,bertsch2000}.
This is an excellent approximation whenever the photon wavelength, $\lambda$, is significantly larger than the spatial extent of the relevant process: at optical and UV frequencies this spans hundreds to thousands of {\AA}ngstroms, so the field is effectively constant across a molecule or unit cell.
Once $\lambda$ becomes comparable to the sample size, unit cell, or electron mean free path~\cite{bonafe2025}, the spatial dependence of $\A(\rr,t)$ can no longer be disregarded.
This is specifically true in the X-ray regime, where the separation of scales collapses.
At $100$\,eV, the lowest energy treated here, $\lambda$ (wavelength) $\approx 124$\,{\AA} and the dipole approximation remains excellent; by $5$\,keV, the highest energy, $\lambda \approx 2.5$\,{\AA} and the dipole approximation has most likely broken down.
The benchmark tests used here fall in between: at the sulfur K-edge of our thiophene benchmark ($2472$\,eV) $\lambda \approx 5$\,{\AA}, comparable to the molecular dimensions; at $1$\,keV in our diamond second harmonic generation (SHG) benchmark $\lambda \approx 12$\,{\AA}, spanning a few unit cells; and at the oxygen K-edge of our CO/Ru benchmark ($\sim$530\,eV) $\lambda \approx 23$\,{\AA}.

Retaining this spatial dependence endows the field with a finite photon momentum $\kgam$ ($|\kgam| = \om/c$ (frequency/speed of light)), and with it access to an entire class of observables that vanish in the dipole limit.
Any process that requires momentum transfer (nondipole asymmetries in photoionization and photoemission~\cite{krassig1995}, X-ray second-harmonic generation~\cite{shwartz2014}, X-ray--optical wave mixing and parametric down-conversion~\cite{eisenberger1971}, and Bragg-geometry nonlinear diffraction~\cite{freund1970}) appears at nonzero reciprocal-space wavevector and is therefore invisible to the macroscopic ($\G=\mathbf{0}$) current that conventional RT-TDDFT reports.
Capturing these effects requires both a spatially varying light--matter coupling and a current diagnostic resolved in reciprocal space.

Traditional RT-TDDFT can in principle imprint some of these signatures on the current density $\jj(t)$, but the mechanism that generates them, a vector potential carrying the spatial structure of the field, is not present in an interaction Hamiltonian limited to the electric dipole represented in either the length or velocity gauge.
Recently, a full minimal-coupling vector potential has been implemented for RT-TDDFT~\cite{bonafe2025}, in a grid-based scheme that also solves self-consistently for the induced fields that demonstrated potential applications with 270\,eV soft X-rays.
However, as $\lambda$ gets smaller and the elements in the system get heavier, complications arise from the treatment of the tightly bound core states where the transitions originate. 
Conventional pseudopotentials which are ubiquitous across periodic/condensed matter simulations freeze and remove these states, and fully grid-based or plane-wave representations can be prohibitively expensive.
Numeric atom-centered orbitals (NAO) are a compact representation that can describe tightly bound core states, and NAO-based RT-TDDFT implementations using specially constructed semi-core pseudopotentials can promote the relevant core electrons to valence~\cite{pemmaraju2018,schacher2026,woodahl2023,helk2021}.
However, such constructions are prone to ghost states and element-specific basis instabilities that constrain which edges can be reliably probed.
An all-electron treatment sidesteps these complications. 
All-electron RT-TDDFT for solids exists in the LAPW framework~\cite{pela2021}, but with the spatially uniform coupling; the compact NAO basis used here reaches the same core states at a fraction of the basis size, treats isolated molecules, slabs, and bulk on the same footing, and, as shown below, accommodates the spatially varying field and its reciprocal-space diagnostics directly.
In this work we present an efficient, all-electron route to the spatially varying field, exploiting its finite photon momentum, in the NAO RT-TDDFT code FHI-aims~\cite{aims}.
Specifically, we:
\begin{enumerate}[label=(\roman*),itemsep=2pt,topsep=4pt]
  \item implement the spatially varying (full minimal coupling) vector
    potential $\A(\rr,t)$ in the time-dependent Kohn--Sham propagation,
    retaining both the paramagnetic $\A\cdot\hat{p}$ (where $\hat{p}$ is the momentum operator) and diamagnetic $\A^2$ responses couplings on an all-electron footing;
  \item recover the density response at the photon wavevector $\kgam$ that the standard $\kk$-diagonal approximation discards using a photon-momentum sideband coupling for the off-diagonal $\kk$-blocks, a first-order expansion in the beyond-dipole coupling;
  \item introduce a reciprocal-space-resolved current diagnostic that expresses
    the paramagnetic and diamagnetic current at nonzero reciprocal-space
    wavevectors $\G$, where the multipole structure and the nonlinear
    harmonic response of arbitrary order reside; and
  \item benchmark the implementation over a wide $100$\,eV to $5$\,keV energy range for several different cases: the molecular
    benzene C K-edge (a direct cross-code validation of the beyond-dipole
    coupling against \citet{bonafe2025}), the thiophene S K-edge
    (the diamagnetic-versus-paramagnetic current crossover of \citet{Rouxel2016}), the second-harmonic response of bulk and slab
    diamond at 1\,keV in the spirit of \citet{shwartz2014}, and
    the CO/Ru(0001) surface O K-edge X-ray absorption of \citet{ostrom2015}.
\end{enumerate}
The remainder of the paper is organized as follows: section~\ref{sec:formalism} develops the formalism, section~\ref{sec:benchmarks} presents the benchmarks, and section~\ref{sec:conclusion} discusses the scope and outlook of the method.

\section{Theoretical Methods}\label{sec:formalism}

\subsection{Minimal-coupling Hamiltonian and the external field}\label{sec:ham_field}

Within Kohn--Sham (KS) RT-TDDFT, the one-electron orbitals $\psi_{n\kk}(\rr,t)$ satisfy~\cite{rungeGross1984,yabana1996}
\begin{equation}
  i\hbar\,\partial_t\,\psi_{n\kk}(\rr,t)
  = \Bigl[\tfrac{1}{2m}\bigl(\hat{p} + \tfrac{e}{c}\sum_{i}\A_{i}(\rr,t) \bigr)^2+ V_\mathrm{KS}[\rho]\Bigr]
    \psi_{n\kk}(\rr,t).
  \label{eq:tdks}
\end{equation}
where $\hbar$ is Planck's constant divided by 2$\pi$, $\partial_t\,\psi_{n\kk}(\rr,t)$ is the time derivative of the KS orbital, $m$ is the mass of an electron, $e$ is the elementary charge (the electron carries charge $-e$), $c$ is the speed of light in a vacuum, $\A_{i}(\rr,t)$ is the magnetic vector potential for pulse $i$ in the Coulomb (velocity) gauge, $\nabla\cdot\A=0$~\cite{cohenTannoudji1997,jackson1999,heras2007}, and $V_\mathrm{KS}[\rho]$ is the KS potential.
Expanding the kinetic energy term for a single pulse,
\begin{equation}
  \tfrac{1}{2m}(\hat{p}+\tfrac{e}{c}\A(\rr,t) )^2
  = \tfrac{\hat{p}^2}{2m}
    + \tfrac{e}{mc}\A(\rr,t) \cdot\hat{p}
    + \tfrac{e^2}{2mc^2}\A^2(\rr,t),
  \label{eq:Hexpand}
\end{equation}
identifies the three contributions: the unperturbed KS kinetic energy, the paramagnetic coupling $\A\cdot\hat{p}$, and the diamagnetic coupling $\A^2$.
The paramagnetic term is responsible for transitions between electronic states driven by the field momentum; the diamagnetic term is a local energy shift proportional to the electron density (the Thomson scattering term).~\cite{blume1985,alsNielsen2011}
Henceforth, we adopt Hartree atomic units ($\hbar = m = e = 1$), retaining the speed of light $c \approx 137$. 

The external field modeled here is a pulsed Gaussian beam in the plane-wave limit.
In the velocity gauge, its vector potential is
\begin{equation}
  \A(\rr,t) = A_0\,f(\rr)\,g(t)\,\cos\!\bigl(\om t - \kgam\cdot\rr + \phi_0\bigr)\,
              \hat{\varepsilon},
  \label{eq:Afield}
\end{equation}
where $A_0$ is the peak vector potential amplitude. 
The spatial envelope is $ f(\rr) = \exp(-r_\perp^2/w_0^2)$ with $r_\perp$ the transverse distance from the beam axis and $w_0$ the beam waist. 
As experiment-like beam waists are on the order of microns, the factor $f(\rr) \approx 1$ for simulations on bulks and slabs that only span a few unit cells and are near the center of the beam, and is thus dropped from the remaining expressions.
It should be noted that a smaller beam waist may be set; however, this makes the field non-transverse ($\nabla\cdot\A \neq 0$), violating the Coulomb-gauge condition assumed here as presented.
The temporal envelope here is Gaussian, $g(t) = \exp\!\bigl(-(t-t_c)^2/(2\sigma_t^2)\bigr)$, with center $t_c$ and width $\sigma_t$.
The photon wavevector is $\kgam = (\om/c)\,\hat{k}$, with propagation direction $\hat{k}$ and magnitude $|\kgam| = \om/c$; the polarization unit vector $\hat{\varepsilon}$ is transverse, $\hat{\varepsilon}\perp\hat{k}$; and $\phi_0$ is the initial carrier phase.
The Fourier decomposition of $\A(\rr,t)$ in spatial wavevectors contains exactly two components for each photon wavevector:
\begin{eqnarray}
  \A(\rr,t) &=& \A(+\kgam,t)\,e^{+i\kgam\cdot\rr} + \A(-\kgam,t)\,e^{-i\kgam\cdot\rr}, \\
  \A(\pm\kgam,t) &=& \frac{A_0\,g(t)}{2}\,e^{\mp i(\om t + \phi_0)}\,\hat{\varepsilon},
  \label{eq:Afourier}
\end{eqnarray}
so the spatial Fourier components are $\A(\pm\kgam,t)$ only, each oscillating at frequency $\mp\om$ (with $A(\pm\kgam,t)=\hat\varepsilon\cdot\A(\pm\kgam,t)$ the scalar amplitude used below).

\subsection{Beyond-dipole coupling in the propagated Hamiltonian}
\label{sec:pA_coupling}

Beyond the unperturbed Hamiltonian $H_0$, the velocity-gauge Hamiltonian of Eq.~\eqref{eq:Hexpand} contains a paramagnetic and a diamagnetic coupling,
\begin{equation}
  H(t) = H_0 + \frac{1}{c}\,\A(\rr,t)\cdot\hat{p}
       + \frac{1}{2c^2}\,\A^2(\rr,t),
  \label{eq:Hcouple}
\end{equation}
both carrying the full spatial dependence of $\A(\rr,t)$.
Using $\hat{p} = -i\nabla$ ($\nabla$ being the spatial gradient) and $\cos(\om t - \kgam\cdot\rr + \phi_0) = \cos(\om t + \phi_0)\cos(\kgam\cdot\rr) + \sin(\om t + \phi_0)\sin(\kgam\cdot\rr)$, the paramagnetic coupling in the atomic orbital (AO) basis ($\mu$, $\nu$), becomes the Hermitian matrix
\begin{equation}
    H^{p\A}_{\mu\nu}(t) = \frac{A_0\,g(t)}{c}\Bigl[           
    \cos(\om t + \phi_0)\,H^{\cos}_{\mu\nu}                                  
    + \sin(\om t + \phi_0)\,H^{\sin}_{\mu\nu}\Bigr], 
  \label{eq:HpA}
\end{equation}   
$H^{\cos} = -\tfrac{i}{2}\bigl(M^{\cos} - (M^{\cos})^\dagger\bigr)$ is an anti-symmetrized matrix (to enforce Hermiticity) bearing the complex prefactor of the momentum operator built from the spatially modulated gradient matrix elements
\begin{equation}
    M^{\cos}_{\mu\nu} = \int \varphi_\mu(\rr)\,\cos(\kgam\cdot\rr)\,
    \bigl(\hat{\varepsilon}\cdot\nabla\varphi_\nu(\rr)\bigr)\,d^3r,
  \label{eq:Mcos}
\end{equation}
and likewise for the analogous $H^{\sin}$ and $M^{\sin}$ with $\cos\to\sin$.
The spatial modulation of $\cos(\kgam\cdot\rr)$ and $\sin(\kgam\cdot\rr)$ is the entire beyond-dipole content; as $\kgam\to\mathbf{0}$, $M^{\cos}$ reduces to the standard velocity-gauge gradient matrix element $\int\varphi_\mu(\rr)(\hat{\varepsilon}\cdot\nabla\varphi_\nu(\rr))\,d^3r$ and $M^{\sin}\to 0$, recovering the spatially uniform (dipole) coupling form of $H^{p\A}_{\mu\nu}(t)$.

The diamagnetic term is a local multiplicative potential, $\A^2(\rr,t) = \tfrac{1}{2}A_0^2\,g^2(t) \bigl[1 + \cos\bigl(2(\om t - \kgam\cdot\rr + \phi_0)\bigr)\bigr]$, that adds a static energy shift together with a $2\om$ component modulated at $2\kgam$.
In the limit $\kgam\to\mathbf{0}$, the diamagnetic coupling reduces to a time-dependent factor scaling an overlap integral, again recovering the spatially uniform (dipole) coupling form~\cite{pemmaraju2018}.
All of these matrices are precomputed once and contracted with the instantaneous field at each step before the time evolution update.
The modulated-gradient integrals $M^{\cos}$ and $M^{\sin}$ are the matrices that source the sideband propagation (Sec.~\ref{sec:sideband_frame}); the reciprocal-space current (Sec.~\ref{sec:Jpara}) is built from the same family of integrals with the complex phase $e^{-i\G\cdot\rr}$.

\subsection{Photon-momentum sidebands of the density matrix}\label{sec:sideband_frame}

\subsubsection{The $\kk$-diagonal approximation}

In a periodic crystal with $N_k$ $\kk$-points, the density matrix has inter-$\kk$ couplings $\rho(\kk,\kk';t)$ for all pairs $\kk,\kk'$ in the Brillouin zone.
The $\kk$-diagonal approximation restricts this to
\begin{equation}
  \rho(\kk,\kk';t) = \rho(\kk;t)\,\delta_{\kk\kk'}.
  \label{eq:kdiag_def}
\end{equation}
Off-diagonal blocks $\rho(\kk,\kk';t)$ with $\kk'\neq\kk$ are set to zero throughout the propagation and $\rho$ is treated as a vector.
This is the standard assumption in all RT-TDDFT codes that do not explicitly construct inter-$\kk$ Hamiltonian matrix elements from the spatially-varying field.
With the $\kk$-diagonal approximation, the real-space electron density is
\begin{equation}
  n(\rr,t) = \sum_{nm,\kk}\rho_{nm}(\kk;t)\,u^*_{n\kk}(\rr)\,u_{m\kk}(\rr),
  \label{eq:n_kdiag}
\end{equation}
which is a sum of products of the lattice-periodic Bloch functions $u_{n\kk}(\rr)$.
As a result, $n(\rr,t)$ is lattice-periodic for all time in the $\kk$-diagonal approximation, regardless of how strongly the field acts.
This has a direct consequence for the density form factor.
The Fourier transform of a lattice-periodic function over an infinite crystal vanishes for any wavevector that is not a reciprocal lattice vector (RLV)~\cite{bloch1929}:
\begin{equation}
 \tfrac{1}{\Omega} \int_{\Omega} n(\rr,t)\,e^{-i\G\cdot\rr}\,d^3r
  \xrightarrow{\Omega\to\infty} 0
  \qquad \text{if } \G \neq \text{RLV},
  \label{eq:nG_zero}
\end{equation}
with $\Omega$ being the volume of the crystal. 
The physical density response to a spatially-varying field necessarily contains a component at $\kgam$ (non-RLV for generic $\om$) driven by the inter-$\kk$ coupling $\rho(\kk,\kk+\kgam;t)$ which is absent in the $\kk$-diagonal approximation.

\subsubsection{Sideband expansion of the density response}
\label{sec:sideband}

A spatially varying periodic field couples Bloch states of different crystal momentum, so the density matrix must be extended by off-diagonal $\kk$-blocks, the photon-momentum sidebands:
\begin{equation}
  \rho_{nm}(\kk,\kk+\ell\kgam;t), \qquad \ell = 0,\pm 1,\pm 2,\ldots,\pm L_\mathrm{max},
  \label{eq:sideband_rho}
\end{equation}
where $L_\mathrm{max}$ is determined by the nonlinear order required.
This expansion has the same space--time harmonic structure as the Floquet--Bloch ladders of temporally driven crystals~\cite{shirley1965,sambe1973,faisal1997} and of traveling-wave-modulated media~\cite{caloz2020,galiffi2022}, which couple $(\kk,\varepsilon)$ to $(\kk+\ell\kgam,\varepsilon+\ell\hbar\om)$; unlike those nonperturbative treatments, we retain only the $\ell=+1$ sideband, sourced at first order in the beyond-dipole coupling from the $\ell=0$ block, so the off-diagonal sector is treated perturbatively while the $\kk$-diagonal propagation remains nonperturbative.
For instance, with SHG where the $2\om$ response is at $2\kgam$, $L_\mathrm{max} = 1$ is sufficient for the leading Freund--Eisenberger diamagnetic channel~\cite{eisenberger1971}.
In general, the reciprocal space representation of the electron density at arbitrary wavevector $\mathbf{q}$ is 
\begin{equation}
  \tilde{n}(\mathbf{q},t) = \frac{1}{\Omega}\int_{\Omega} n(\rr,t)\,
    e^{-i\mathbf{q}\cdot\rr}\,d^3r,
\end{equation}
and when the off-diagonal density response at $\kgam$ is factored into Eq.~\eqref{eq:n_kdiag} and taken to reciprocal space the resulting expression is
\begin{equation}
  \tilde{n}(\kgam,t) = \frac{1}{\Omega}\sum_{\kk,\ell}\mathrm{Tr}[\rho(\kk+\ell\kgam,\kk;t)\,F^{\G_\ell}] + \text{h.c.}
  \label{eq:nkgam_full}
\end{equation}
where \text{h.c.} indicates the Hermitian conjugate and with $\rho$ including the occupations $f_{n\kk}$ as in Eq.~\eqref{eq:rho_offdiag}. 
This response is zero in the $\kk$-diagonal approximation but nonzero in the full treatment.
$F^{\G_{\ell}}$ is the structure-factor matrix for reciprocal space vector $\G_{\ell} = \ell\kgam$,
\begin{equation}
  F^{\G_{\ell}}_{\mu\nu}
  = \int \varphi_\mu(\rr)\,e^{-i\G_{\ell}\cdot\rr}\,\varphi_\nu(\rr)\,d^3r,
  \label{eq:FGmat}
\end{equation}
which is time-independent and $\kk$-independent.
$F^{\G_{\ell}}_{\mu\nu}$ is in the AO basis and is precomputed at the outset of the simulation.
The extraction phase $e^{-i\G_{\ell}\cdot\rr}$ between the atom-centered states at $\kk$ and $\kk+\ell\kgam$ in the single-cell ($R = 0$) representation is not unity because the density Fourier component is taken at $\kgam$ rather than at a RLV.

To accomplish time propagation, the $\kk$-diagonal component is propagated using the standard FHI-aims RT-TDDFT propagator, while we implement a separate scheme for the off-diagonal inter-$\kk$ components.
The basic algorithm is that the $\kk$-diagonal representation is propagated in the standard approach yielding a set of time dependent Bloch function expansion coefficients, $\mathbf{C}^\mathrm{diag}(t)$.
At each timestep, the inter-$\kk$ off-diagonal contribution is propagated after the diagonal contribution has been propagated and stored in a separate set of expansion coefficients, $\mathbf{C}^\mathrm{off}(t)$.
These sets of coefficients are not combined during the simulation, and the $\kk$-diagonal propagation never sees the inter-$\kk$ contribution.
The inter-$\kk$ density matrix in the AO basis, which is what the structure matrices $F^{\G_{\ell}}_{\mu\nu}$ are contracted with, is built from the Bloch expansion coefficients $C_{\mu n}(\kk,t)$ of the occupied states,
\begin{equation}
  \rho_{\mu\nu}(\kk+\kgam,\kk;t)
  = \sum_{n} f_{n\kk}\,C_{\mu n}(\kk+\kgam,t)\,C^{*}_{\nu n}(\kk,t)
  \;\approx\; \sum_{n} f_{n\kk}\,\mathbf{C}^\mathrm{off}_{\mu n}(\kk,t)\,
    \bigl[\mathbf{C}^\mathrm{diag}_{\nu n}(\kk,t)\bigr]^{*},
  \label{eq:rho_offdiag}
\end{equation}
that is, $\rho(\kk+\kgam,\kk;t) \approx \mathbf{C}^\mathrm{off}(\kk,t)\,f_{\kk}\,[\mathbf{C}^\mathrm{diag}(\kk,t)]^{\dagger}$, where $\mathbf{C}^\mathrm{off}(\kk,t)$ holds the amplitude at $\kk+\kgam$ driven from occupied Bloch state $n$ at $\kk$, and the reverse block is its Hermitian conjugate, $\rho(\kk,\kk+\kgam;t) = [\rho(\kk+\kgam,\kk;t)]^{\dagger}$.
Contracting with $F^{\G_1}$ and summing over $\kk$ gives the density Fourier component of Eq.~\eqref{eq:nkgam_full}, evaluated in practice as Eq.~\eqref{eq:sideband_nG1} below.
The time evolution of the inter-$\kk$ expansion coefficients is performed here using the Crank--Nicolson propagator~\cite{castro2004,yabana2012}:
\begin{equation}
\begin{split}
  \bigl(S_\kk + \tfrac{i\,\mathrm{d}t}{2}\,H^\mathrm{mid}_\kk\bigr)\,
    \mathbf{C}^\mathrm{off}(\kk,t+dt)
  = \bigl(S_\kk - \tfrac{i\,\mathrm{d}t}{2}\,H^\mathrm{mid}_\kk\bigr)\,
    \mathbf{C}^\mathrm{off}(\kk,t) \\
  - \mathrm{d}t\,g_\mathrm{mid}
    \bigl[\cos_\mathrm{mid}\,M^\mathrm{raw}_\mathrm{cos}
         + \sin_\mathrm{mid}\,M^\mathrm{raw}_\mathrm{sin}\bigr]
    \mathbf{C}^\mathrm{diag}(\kk,t),
  \label{eq:sideband_cn}
\end{split}
\end{equation}
where $S_\kk$ is the overlap matrix, $H^\mathrm{mid}_\kk = H_\kk(t+\mathrm{d}t/2)$ is the Crank--Nicolson midpoint Hamiltonian, $g_\mathrm{mid} = g(t+\mathrm{d}t/2)$ and $\cos[\sin]_\mathrm{mid} = \cos[\sin](\om(t+\mathrm{d}t/2)+\phi_0)$ are the temporal envelope and carrier evaluated at the step midpoint, and $M^\mathrm{raw}_{\mathrm{cos[sin]}} = (A_0/c)\int\varphi_i\,(\hat\varepsilon\cdot\nabla\varphi_j)\, \cos[\sin](\kgam\cdot\rr)\,\mathrm{d}^3r$ are the raw (non-antisymmetrized) $p\cdot\A$ integration matrices of Eq.~\eqref{eq:Mcos}, scaled by the field-amplitude prefactor $A_0/c$ of the coupling in Eq.~\eqref{eq:HpA}.
The displayed real coefficient $-\mathrm{d}t\,g_\mathrm{mid}[\cos_\mathrm{mid}M^\mathrm{raw}_\mathrm{cos} + \ldots]$ is complete: the factor $-i$ from $\hat p = -i\nabla$ in the velocity-gauge coupling cancels the $-i$ of the Crank--Nicolson rearrangement, so no explicit $i$ remains.
The off-diagonal propagation uses the raw rather than the antisymmetrized $M$ matrices for two reasons.
First, for a transverse field ($\hat{\varepsilon}\perp\kgam$, $\nabla\cdot\A = 0$) the operator $\A\cdot\hat{p}$ is Hermitian analytically, so on a complete basis with exact quadrature the one-sided integral $M^\mathrm{raw}$ and its antisymmetrized form coincide; the antisymmetrization in Eq.~\eqref{eq:HpA} only enforces Hermiticity of the $\kk$-diagonal block against finite-basis and quadrature error.
Second, the $\kk\to\kk+\kgam$ block of a Hermitian operator is not itself a Hermitian matrix: its adjoint is the $\kk+\kgam\to\kk$ ($\ell = -1$) block, so antisymmetrizing within the $\ell = +1$ block would mix in the wrong matrix element, and the one-sided integral is the correct block.
Note also that in the single-cell ($R = 0$) representation the Bloch phase $e^{\pm i\kgam\cdot\rr}$ is carried by the modulated integrals rather than by the basis functions, so $\mathbf{C}^\mathrm{off}$ holds the full real first-order correction to the occupied orbitals driven by the real carrier $\cos(\om t - \kgam\cdot\rr + \phi_0)$; the assignment to the $+\kgam$ sideband is made at readout, by the structure matrix $F^{\G_1}$ in Eq.~\eqref{eq:sideband_nG1}.
The key approximations built into Eq.~\eqref{eq:sideband_cn} are:
\begin{enumerate}
\item $H(\kk+\kgam) \approx H(\kk)$: the free-propagation operator on the left-hand side uses the same-$\kk$ Hamiltonian and overlap. 
In the single-cell representation, $H(\kk)$ and $H(\kk+\kgam)$ differ only through the Bloch phases $e^{i\kgam\cdot\R}$ of the inter-cell matrix elements, so the approximation neglects the band detuning $\varepsilon_n(\kk+\kgam) - \varepsilon_n(\kk)$ in the free evolution of the $\ell=+1$ amplitude between its creation by the source term and its readout; the resulting error is a dephasing of relative order $|\varepsilon_n(\kk+\kgam) - \varepsilon_n(\kk)|\,\tau/\hbar$ over the source-to-readout interval $\tau$, bounded by the pulse duration. 
For the molecular benchmarks (Secs.~\ref{sec:benzene} and~\ref{sec:thiophene}) there are no Bloch phases and the approximation is exact. 
For the diamond benchmark (Sec.~\ref{sec:diamond}) $|\kgam| = 0.506$\,{\AA}$^{-1}$ is one fifth of the in-plane reciprocal-lattice vector of the $(001)$ cell, so the detuning is not negligible a priori; its effect is confined to the $\ell = +1$ density correction, i.e.\ to the diamagnetic $J^\mathrm{dia}(\G_2,2\om)$ channel of Eq.~\eqref{eq:JdiaG2} and the $n(\G_2)$ term of Eq.~\eqref{eq:JdiaG1}, since the $\kk$-diagonal propagation never sees $\mathbf{C}^\mathrm{off}$ and the $\G = \mathbf{0}$ forward-SHG, dipole, and absorbed-energy observables are unaffected.
\item $\mathbf{C}^\mathrm{diag}(\kk,t+\tfrac{dt}{2}) \approx \mathbf{C}^\mathrm{diag}(\kk,t)$:
a consistent Crank--Nicolson treatment would evaluate the source at the step midpoint; instead the source term uses the $\kk$-diagonal state at the start of the step, $\mathbf{C}^\mathrm{diag}(\kk,t)$, which is the state available before the diagonal update (first-order operator splitting between the diagonal and off-diagonal propagations, with an error of order $\mathrm{d}t$ in the source that is already first order in the beyond-dipole coupling).
\item Only the $\ell=+1$ sideband is tracked ($L_\mathrm{max} = 1$), with $\ell = -1$ following from the Hermitian conjugate of Eq.~\eqref{eq:rho_offdiag}. 
This is sufficient for the leading diamagnetic SHG channel, the Freund--Eisenberger term $A(+\kgam,t)\,n(\G_1,t)$ of Eq.~\eqref{eq:JdiaG2}, which is first order in the beyond-dipole coupling.
Higher channels also contribute at $2\om$: the paramagnetic $\kk\to\kk+2\kgam$ pathway ($\ell = 2$) and the $n(\G_3,t)$ term of Eq.~\eqref{eq:JdiaG2} ($\ell = 3$) enter at second and third order in the coupling and are not propagated here; $n(\G_3,t)$ in Eq.~\eqref{eq:JdiaG2} is accordingly evaluated from the $\kk$-diagonal density alone, without a sideband coupling correction.
\end{enumerate}

The off-diagonal coefficients then contribute to the update of $\tilde{n}(\kgam,t)$ from the previous timestep via (retaining only the on-site $R=0$ term of the Born--von K\'arm\'an lattice sum, i.e.\ the single-unit-cell approximation):
\begin{equation}
\delta\tilde{n}(\kgam,t) = 2\,\mathrm{Re}
  \sum_{\kk,s} w_s \sum_{n} f_{n\kk}
  \sum_{\mu\nu}\bigl[\mathbf{C}^\mathrm{diag}_{\mu n}(\kk,t)\bigr]^{*}\,
    F^{\G_1}_{\mu\nu}(s)\,\mathbf{C}^\mathrm{off}_{\nu n}(\kk,t),
  \label{eq:sideband_nG1}
\end{equation}
where the $F^{\G_1}_{\mu\nu}$ matrix (and all other non-expansion coefficient AO-basis objects) is represented on a real-space Becke--Lebedev quadrature grid~\cite{becke1988,lebedev1999,havu2009} spanning the simulation volume, so the integral is performed discretely by a summation over all grid points $s$ with weight $w_s$. 
$f_{n\kk}$ in Eq.~\eqref{eq:sideband_nG1} is the Fermi occupation number of Bloch function $n$ at k-point $\kk$ as determined during the $\kk$-diagonal propagation.
This correction is added to $\tilde{n}(\G_1,t)$ at each output step before the current density evaluation enabling determination of the Freund--Eisenberger SHG channel.

\subsection{The reciprocal-space-resolved current density}

The gauge-invariant physical current density is the primary metric for the response of the system to the applied field, and it has both a paramagnetic and diamagnetic component: 
\begin{gather}
  \jj(\rr,t) = \jj^\mathrm{para}(\rr,t) + \jj^\mathrm{dia}(\rr,t)
  \label{eq:Jtotal}\\
  \jj^\mathrm{para}(\rr,t)
  = -\sum_{n\kk}f_{n\kk}\,\mathrm{Im}\!\bigl[\psi_{n\kk}^*\nabla\psi_{n\kk}\bigr]
   = -\frac{1}{2i}\sum_{n\kk}f_{n\kk}
    \bigl[\psi_{n\kk}^*\nabla\psi_{n\kk} - \psi_{n\kk}\nabla\psi_{n\kk}^*\bigr],
  \label{eq:jpara_real}\\
  \jj^\mathrm{dia}(\rr,t)
  = -\frac{1}{c}\,n(\rr,t)\,\A(\rr,t).
  \label{eq:jdia_real}
\end{gather}
Here $n(\rr,t) = \sum_{n\kk}f_{n\kk}|\psi_{n\kk}(\rr,t)|^2$ is the time-dependent electron density.
The correspondence with the minimal-coupling Hamiltonian of Eq.~\eqref{eq:Hexpand} (ie. $\hat{H}_\text{int}=-\int d\rr\, \jj(\rr,t)\cdot\A(\rr,t)$) is direct: $\jj^\mathrm{para}$ constitutes the $\A\cdot\hat{p}$ term (paramagnetic, requires electronic transitions) and $\jj^\mathrm{dia}$ constitutes the $\A^2$ term (diamagnetic, driven by the instantaneous density).
This decomposition is identical to the one used by \citet{Rouxel2016} in their nonlinear X-ray response formalism, wherein $\jj^\mathrm{para}$ dominates near resonance (photon energy close to an electronic transition) while $\jj^\mathrm{dia}$ dominates off resonance, far from any absorption edge (e.g. at $1500$\,eV in sulfur which is between the L edge at $230$\,eV and the K-edge at $2472$\,eV).~\cite{henke1993}

In a periodic crystal, the spatial Fourier transform of the current density at reciprocal space vector $\G$ is
\begin{equation}
  \jj(\G,t) = \frac{1}{\Omega}\int_{\Omega} \jj(\rr,t)\,e^{-i\G\cdot\rr}\,d^3r.
  \label{eq:JGdef}
\end{equation}
The standard $\kk$-diagonal macroscopic current, $\jj(t) = \jj(\G=\mathbf{0},t)$, is blind to spatial structure at non-zero $\G$.
For a field with wavevector $\kgam$, the linear optical response oscillates at $e^{\pm i\kgam\cdot\rr}$ and the second harmonic response at $e^{\pm 2i\kgam\cdot\rr}$; both are invisible to the $\G=\mathbf{0}$ channel.
Thus, higher $\G$-resolved output is essential for observing multipole and Bragg-geometry responses for nonlinear X-ray phenomena such as SHG.
To obtain the inter-$\kk$ contributions to the current density, two $\G$-vectors beyond the diagonal component are tracked:
\begin{equation}
  \G_1 = \frac{\om}{c}\,\hat{k}  = \kgam , \qquad
  \G_2 = \frac{2\om}{c}\,\hat{k} = 2\G_1 = 2\kgam.
  \label{eq:Gvecs}
\end{equation}
A third vector $\G_3 = 3\G_1$ is precomputed for use in the diamagnetic convolution (Sec.~\ref{sec:jdia_conv}).

\subsubsection{Paramagnetic component}
\label{sec:Jpara}

Inserting the time-dependent density matrix represented in the AO basis, $\rho_{\mu\nu}(\kk,t)$, into Eq.~\eqref{eq:jpara_real} yields
\begin{equation}
  J^\mathrm{para}_\alpha(\G_n,t)
  = -\frac{1}{\Omega}\,\mathrm{Tr}\bigl[\rho(t)\,P^{\G_n}_\alpha\bigr],
  \label{eq:JparaG}
\end{equation}
where the precomputed modulated gradient matrix in the AO basis is
\begin{equation}
  P^{\G_n}_{\alpha,\mu\nu}
  = \int \varphi_\mu(\rr)\,e^{-i\G_n\cdot\rr}\,\partial_\alpha\varphi_\nu(\rr)\,d^3r.
  \label{eq:PGmat}
\end{equation}
In practice, that matrix and all others in the AO basis are computed on a discrete real-space Becke--Lebedev quadrature grid~\cite{becke1988,lebedev1999,havu2009} spanning the simulation volume, $\Omega$:
\begin{equation}
  P^{\G_n}_{\alpha,\mu\nu}
  = \sum_s \varphi_\mu(\rr_s)\,e^{-i\G_n\cdot\rr_s}\,\partial_\alpha\varphi_\nu(\rr_s)\,
    w_s,
  \label{eq:PGmat_grid}
\end{equation}
where $\rr_s$ and $w_s$ are the position and weight of grid point $s$, respectively.
The matrix $P^{\G_n}_{\alpha,\mu\nu}$ is $\kk$-independent: the atomic basis functions do not carry Bloch phases, so $P^{\G_n}_{\alpha,\mu\nu}$ is computed once before propagation begins and stored in memory.
Equation~\eqref{eq:JparaG} is evaluated directly as a complex trace.
Because $P^{\G_n}_{\alpha,\mu\nu}$ carries the one-sided gradient, Eq.~\eqref{eq:JparaG} omits the explicit factor $1/(2i)$ and the $\psi\nabla\psi^*$ term of Eq.~\eqref{eq:jpara_real}; the two are related exactly.
Writing $T_\alpha(\G,t) = \tfrac{1}{\Omega}\mathrm{Tr}[\rho(t)\,P^{\G}_\alpha] = \tfrac{1}{\Omega}\sum_{n\kk}f_{n\kk}\int\psi^*_{n\kk}\,e^{-i\G\cdot\rr}\,\partial_\alpha\psi_{n\kk}\,d^3r$ for the one-sided trace, the Fourier transform of the second term of Eq.~\eqref{eq:jpara_real} is $T_\alpha(-\G,t)^*$, and integrating by parts (the boundary term vanishes for lattice-periodic or localized orbitals) gives the identity
\begin{equation}
  T_\alpha(-\G,t)^* = -\,T_\alpha(\G,t) + i\,G_\alpha\,n(\G,t),
  \label{eq:onesided_identity}
\end{equation}
with $n(\G,t)$ the density form factor.
The physical Fourier component of the paramagnetic current is therefore
\begin{equation}
  j^\mathrm{para}_\alpha(\G,t)
  = -\frac{1}{2i}\bigl[T_\alpha(\G,t) - T_\alpha(-\G,t)^*\bigr]
  = i\,T_\alpha(\G,t) + \tfrac{1}{2}\,G_\alpha\,n(\G,t).
  \label{eq:jpara_from_trace}
\end{equation}
For $\G=\mathbf{0}$ the identity makes $T(\mathbf{0})$ purely imaginary and Eq.~\eqref{eq:jpara_from_trace} reduces to $-\mathrm{Im}\,T(\mathbf{0})$, the standard macroscopic current.
For $\G\neq\mathbf{0}$ the second term is longitudinal (parallel to $\G$) and drops out of the polarization-projected component whenever $\hat{\varepsilon}\perp\G$, which holds for every $\G_n = n\kgam$ of Eq.~\eqref{eq:Gvecs} with a transverse field, leaving $\hat{\varepsilon}\cdot\jj^\mathrm{para}(\G_n,t) = -i\,\hat{\varepsilon}\cdot\mathbf{J}^\mathrm{para}(\G_n,t)$ with $\mathbf{J}^\mathrm{para}$ the recorded trace of Eq.~\eqref{eq:JparaG}.
The recorded and physical currents thus differ exactly by the unit-modulus factor $-i$, which cancels in the magnitude observables of Sec.~\ref{sec:Jobs}.

\subsubsection{Diamagnetic component}
\label{sec:jdia_conv}

The inter-$\kk$ diamagnetic current at $\G_n$ requires the Fourier-space convolution of the field and density.
From Eq.~\eqref{eq:jdia_real},
\begin{equation}
  J^\mathrm{dia}_\alpha(\G_n,t)
  = -\frac{1}{c\Omega}\sum_{\G'} A_\alpha(\G',t)\,n(\G_n - \G',t),
  \label{eq:JdiaG_conv}
\end{equation}
where $n(\G,t) = \mathrm{Tr}[\rho(t)\,F^{\G}]$ is the time-dependent density form factor.
With the two-component field of Eq.~\eqref{eq:Afourier}, only $\G' = \pm\kgam$ contribute:
\begin{align}
  J^\mathrm{dia}_\alpha(\G_1,t)
  &= -\frac{\hat{\varepsilon}_\alpha}{c\Omega}\bigl[
    A(+\kgam,t)\,n(\G_1-\kgam,t) + A(-\kgam,t)\,n(\G_1+\kgam,t)
    \bigr] \notag\\
  &= -\frac{\hat{\varepsilon}_\alpha}{c\Omega}\bigl[
    A(+\kgam,t)\,N_e + A(-\kgam,t)\,n(\G_2,t)
    \bigr],
  \label{eq:JdiaG1}\\[4pt]
  J^\mathrm{dia}_\alpha(\G_2,t)
  &= -\frac{\hat{\varepsilon}_\alpha}{c\Omega}\bigl[
    A(+\kgam,t)\,n(\G_1,t) + A(-\kgam,t)\,n(\G_3,t)
    \bigr],
  \label{eq:JdiaG2}
\end{align}
using $\G_1 - \kgam = \mathbf{0}$ (so $n(\mathbf{0}) = N_e$, exact), $\G_1 + \kgam = \G_2$, $\G_2 - \kgam = \G_1$, and $\G_2 + \kgam = \G_3$.

Equation~\eqref{eq:JdiaG2} is the diamagnetic second-harmonic channel: the field component $A(+\kgam,t)\propto e^{-i\om t}$ beats against the density modulation $n(\G_1,t)\propto e^{-i\om t}$ to source a $2\om$ current at $\G_2$, ie. the Freund--Eisenberger cold-plasma mechanism~\cite{eisenberger1971}.
The density form factors that drive these expressions, $n(\G_1,t)$ in Eq.~\eqref{eq:JdiaG2} and $n(\G_2,t)$ in Eq.~\eqref{eq:JdiaG1}, are evaluated at the wavevectors $\kgam$ and $2\kgam$, which are not reciprocal lattice vectors for generic $\om$.
As shown in Sec.~\ref{sec:sideband_frame}, these components vanish in the standard $\kk$-diagonal approximation; the diamagnetic G-resolved current beyond the exact $n(\mathbf{0}) = N_e$ contribution is therefore recovered only through the sideband coupling.

\subsubsection{Extracting observables}
\label{sec:Jobs}

At each output step the polarization-projected components $J^\mathrm{para}_{\hat{\varepsilon}}(\G_n,t) = \hat{\varepsilon}\cdot\jj^\mathrm{para}(\G_n,t)$ and $J^\mathrm{dia}_{\hat{\varepsilon}}(\G_n,t)$ are recorded for $\G_1$ and $\G_2$, together with their $-\G$ partners (related by $J(-\G,t) = J(\G,t)^*$ for a real current density) as well as the $\kk$-diagonal contribution.
A discrete Fourier transform in time yields the spectra $J(\G_n,\om)$, from which the physical observables follow: the power spectrum $|J(\G_n,\om)|$, the second-harmonic signal $|J^\mathrm{dia}(\G_2,2\om)|$ (yielding an effective second-order susceptibility $|\chi^{(2)}| \propto |J^\mathrm{dia}(\G_2,2\om)|/|A(\om)|^2$ after normalization by the driving field), and the channel ratio
\begin{equation}
  R(\om) = \frac{|J^\mathrm{dia}(\G_1,\om)|}{|J^\mathrm{para}(\G_1,\om)|},
  \label{eq:Rratio}
\end{equation}
which quantifies the relative weight of the diamagnetic and paramagnetic channels in the classification of \citet{Rouxel2016}.
All of these observables depend only on spectral magnitudes.

The beyond-dipole test of Sec.~\ref{sec:benzene} instead uses the real-space electronic dipole, evaluated directly from the propagated density as the expectation value of the position operator,
\begin{equation}
  \mu_\alpha(t) = -\,\mathrm{Tr}\bigl[\rho(t)\,D_\alpha\bigr], \qquad
  D_{\alpha,\mu\nu} = \int \varphi_\mu(\rr)\,r_\alpha\,\varphi_\nu(\rr)\,d^3r,
  \label{eq:dipole}
\end{equation}
with $D_\alpha$ the position-operator matrix in the AO basis, integrated once on the same Becke--Lebedev grid as $P^{\G_n}$, and $\rho(t)$ the density matrix of Eq.~\eqref{eq:JparaG}.
This is a length-form ($\langle r\rangle$) observable taken straight from the density rather than the time integral of the velocity-gauge current; the two are equivalent through $\dot{\mu}_\alpha = -J_\alpha$, but the direct form avoids accumulating quadrature drift.~\cite{schelter2018,li2020} 
Since $D_\alpha$ is the uniform position operator, $\mu_\alpha(t)$ is the $\G=\mathbf{0}$ projection of the polarization, so all beyond-dipole content enters through the spatial structure of $\rho(t)$ and the difference between the induced dipole calculated via full minimal coupling (FMC) and electric dipole (ED) approximation $\mu_\alpha^\mathrm{FMC}(t) - \mu_\alpha^\mathrm{ED}(t)$ isolates the response generated by the spatial phase of the field.

For an extended, periodic system the real-space dipole of Eq.~\eqref{eq:dipole} is ill-defined, so the metallic-slab benchmark of Sec.~\ref{sec:ostrom} uses a third observable, the energy absorbed from the pulse, which is a scalar and requires no choice of origin.
It is the drift in the total electronic energy across the pulse with frequency $\om_0$, \cite{yabana2012}
\begin{equation}
  \dEabs(\om_0) = E_\mathrm{el}(t_f; \om_0) - E_\mathrm{el}(0),
  \label{eq:Eabs}
\end{equation}
read from the post-pulse plateau of the total Kohn--Sham energy that the propagator already evaluates at each step, with $t_f$ being the time when the envelope has switched off and the nuclei clamped so that the drift is purely electronic excitation.
By the work--energy theorem, $\dEabs$ is the work done by the field on the electrons, and in linear response $\dEabs \propto \int |\tilde{E}(\om)|^2\,\mathrm{Re}\{\sigma(\om)\}\,d\om$,  the absorptive part of the conductivity weighted by the pulse spectrum \cite{yabana2012}; for a narrowband carrier this is dominated by $\mathrm{Re}\{\sigma(\om_0)\}$, so scanning the carrier traces the absorption lineshape and the difference $\dEabs^\mathrm{FMC} - \dEabs^\mathrm{ED}$ isolates the beyond-dipole contribution to the absorption.
Unlike the current and dipole observables, $\dEabs$ is a single number per propagation rather than a spectrum, which is why the O K-edge lineshape of Sec.~\ref{sec:ostrom} is reconstructed pointwise from a scan of carriers.

\section{Benchmarks}\label{sec:benchmarks}

We validate the implementation with calculations spanning the XUV to the hard-X-ray regime.
The first two examples are molecular tests against established references, validating the beyond-dipole light--matter coupling and the current decomposition, respectively; the remaining examples probe the reciprocal-space and the photon-momentum sideband coupling in periodic systems.

\subsection{Benzene at the carbon K-edge: beyond-dipole response against full minimal coupling}
\label{sec:benzene}

We begin with a direct validation of the beyond-dipole light--matter coupling itself against an independent full-minimal-coupling reference.
\citet{bonafe2025} introduced a self-consistent Maxwell--TDDFT framework in which the electrons feel the full spatial profile of $\A(\rr,t)$ rather than its dipole (long-wavelength) limit $\A(t)$, the propagated TDDFT current is a source for the microscopic Maxwell equations, and the resulting fields act back onto the electrons.
The example from their work that is most relevant here is benzene driven by an XUV pulse at the carbon K-edge, for which they report the induced dipole $\mu_x(t)$ in the ED and with FMC, and identify a beyond-dipole correction $\mu_x^\mathrm{FMC}(t) - \mu_x^\mathrm{ED}(t)$ of $\sim 5\%$ of the dipole response, consistent with the geometric estimate $(|\kgam|\,r)^2/2$~\cite{bonafe2025}.

This is a direct, quantitative cross-code check of the beyond-dipole $\A\cdot\hat{p}$ coupling between two all electron codes with significant differences in implementation.
A numeric atom-centered basis is used here, whereas a real-space grid is used in Octopus~\cite{tancognedejean2020}. 
The system is a single 12-atom $D_{6h}$ benzene molecule in the $xy$ plane (C--C $= 1.399$\,{\AA}, C--H $= 1.084$\,{\AA}), treated as an isolated molecule.
Following Bonaf\'e's geometry, the field propagates in-plane along $\hat{k} = (0,1,0)$ with polarization $\hat{\varepsilon} = (1,0,0)$, so the spatial phase $e^{i(\om/c)y}$ varies across the ring.
Under this in-plane geometry, the out-of-plane C\,$1s\to\pi^*$ transition is dipole-forbidden and the $x$-polarized dipole response is carried by the in-plane $\sigma^*$ channels, so the FMC--ED difference isolates the beyond-dipole oscillator-strength transfer rather than competing with a large dipole-allowed background.
Simulations were performed using the PW92 LDA functional~\cite{perdewWang1992}, velocity-gauge Crank--Nicolson propagation, a perturbative amplitude $A_0 = 10^{-4}$\,a.u., and a Gaussian pulse with $\sigma_t = 0.15$\,fs (spectral FWHM $\approx 10$\,eV) centered at $t_c = 0.25$\,fs propagated for $T = 1.0$\,fs with $\Delta t = 5\times10^{-4}$\,fs.
We scan nine carrier energies from $263$ to $303$\,eV in $5$\,eV steps, each as an ED/FMC pair, bracketing the LDA C\,$1s\to\pi^*$ resonance which at this level of theory lies at $265$\,eV (the LUMO$-$C\,$1s$ Kohn--Sham gap).
As in the other benchmarks, this is red-shifted from the experimental benzene C\,$1s\to\pi^*$ resonance at $285.0$\,eV~\cite{rennie2000} by $\sim 20$\,eV due to the typical DFT band gap error~\cite{perdewZunger1981}.

\begin{figure}[htbp]
  \centering
  \includegraphics[width=0.6\textwidth]{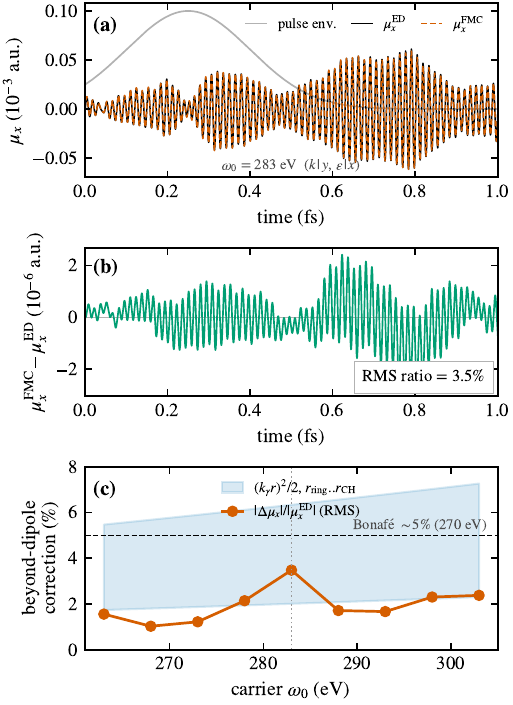}
  \caption{Beyond dipole-dipole response of benzene at the carbon K-edge, the analogue of \citet{bonafe2025}, in the in-plane geometry ($\hat{k}\,\|\,\hat{y}$, $\hat{\varepsilon}\,\|\,\hat{x}$).
  \textbf{(a)} Induced dipole $\mu_x(t)$ at $\om_0 = 283$\,eV in the dipole approximation (ED) and with full minimal coupling (FMC); the two overlap at this scale, with the pulse envelope shown for reference.
  \textbf{(b)} The beyond-dipole difference $\mu_x^\mathrm{FMC} - \mu_x^\mathrm{ED}$, with RMS ratio $3.5\%$ of the dipole response.
  \textbf{(c)} The RMS ratio $|\Delta\mu_x|/|\mu_x^\mathrm{ED}|$ across the $263$--$303$\,eV scan against the geometric $(|\kgam|\,r)^2/2$ band spanned by the ring and C--H radii; the correction is $1$--$3.5\%$, comparable to the $\sim 5\%$ reported by Bonaf\'e at $270$\,eV (dashed).
  Energies are on the LDA scale, with the C\,$1s\to\pi^*$ resonance at $265$\,eV.}
  \label{fig:benzene}
\end{figure}

The observable in this case is the beyond-dipole correction to the induced dipole, $\Delta\mu_x(t) = \mu_x^\mathrm{FMC}(t) - \mu_x^\mathrm{ED}(t)$, the direct analogue of Bonaf\'e  et al.
We quote its root-mean-square magnitude relative to the dipole response, $|\Delta\mu_x|/|\mu_x^\mathrm{ED}|$, evaluated over the post-onset interval.
The driven ($x$) component is used because it is symmetry-clean, and the FMC$-$ED difference cancels the small grid-quadrature noise common to both runs.
Results are shown in Figure~\ref{fig:benzene}.
The ED and FMC dipoles overlap at the scale of the response (panel a); their difference (panel b) is the pure beyond-dipole content, with an RMS magnitude of the FMC correction of $3.5\%$ of the dipole response at $\om_0 = 283$\,eV. 
Across the full energy scan, the correction is $1$--$3.5\%$ (panel c), peaking at $283$\,eV where $\mu_x^\mathrm{ED}$ passes through a minimum and the relative beyond-dipole weight is therefore largest.

The natural scale for this correction is set by the spatial phase of the field.
Expanding it, $e^{i\kgam\cdot\rr} = 1 + i\,\kgam\cdot\rr - \tfrac{1}{2}(\kgam\cdot\rr)^2 + \cdots$, separates the dipole limit (the unit term) from the beyond-dipole corrections, with the small parameter $|\kgam|\,r = (\om/c)\,r$, the carrier phase, accumulated across the relevant length scale $r$.
At $\om = 283$\,eV, $|\kgam| = 0.144$\,{\AA}$^{-1}$ and the phase accumulated across the molecule is therefore $|\kgam|\,r_\mathrm{CH} \approx 0.36$ with $r_\mathrm{CH} \approx 2.5$\,{\AA} being the planar radius of the ring including the hydrogens --- the largest geometric scale of the molecule.
The two leading corrections enter the driven dipole differently.
Expanding the carrier's spatial phase, $\cos(\om t - \kgam\cdot\rr) = \cos(\om t)\cos(\kgam\cdot\rr) + \sin(\om t)\sin(\kgam\cdot\rr)$, the leading term $\cos(\om t)\cos(\kgam\cdot\rr)$ is even in $\rr$ and in phase with the carrier (the uniform dipole drive), while the first-order term $\sin(\om t)\sin(\kgam\cdot\rr) \approx (\kgam\cdot\rr)\sin(\om t)$ is odd in $\rr$ and $90^{\circ}$ out of phase with respect to time with it. Being odd and out of phase, this term carries the higher order content of the field (electric-quadripole and magnetic-dipole), so for the centrosymmetric benzene ring it drives a transverse, out of phase response rather than in phase dipole along $\hat\varepsilon$. These channels carry the orbital angular momentum signal that Bonaf\'e et al.\ use as their resonant beyond-dipole discriminator.

The leading correction to the in-phase $\mu_x$ is therefore the even, second-order term $\cos(\kgam\cdot\rr) \approx 1 - \tfrac{1}{2}(\kgam\cdot\rr)^2$, which reduces the coherent $\A\cdot\hat{p}$ coupling as the carrier phase slips across the molecule.
The relative correction is thus of order $\tfrac{1}{2}(|\kgam|\,r)^2$, with $r$ the characteristic extent of the C\,$1s$ transition density; bounding $r$ between the benzene ring radius ($r_\mathrm{ring} \approx 1.4$\,{\AA}) and the carbon--hydrogen radius of the ring $r_\mathrm{CH}$ defined above gives the blue band in Fig.~\ref{fig:benzene}(c), $2.0\%$ to $6.4\%$ at $283$\,eV.
Our measured $1$--$3.5\%$ falls within this band, tracking its lower (ring-radius) edge, and is comparable to the reported $\sim 5\%$ at $270$\,eV, which corresponds to the larger C--H scale.
This implementation does not take into account the self-consistent back-action of the induced field, and therefore the magnitude of the correction is expected to be smaller in comparison.
Given the large differences in implementations (ie. basis set quality and representation) and the lack of self-consistent back action, the agreement between the two methods is quite strong.

\FloatBarrier
\subsection{Thiophene across the sulfur and carbon edges: diamagnetic
versus paramagnetic response}
\label{sec:thiophene}

Our second benchmark is thiophene (C$_4$H$_4$S, $N_e = 44$ electrons), the molecule used in the numerical demonstrations of \citet{Rouxel2016}.
Whereas Rouxel et al.\ evaluate the current ($\jj^\mathrm{para}\cdot\A$, paramagnetic) and charge-density ($\sigma\A^2$, diamagnetic) contributions perturbatively from a sum over states, we extract them non-perturbatively from a single $G$-resolved real-time propagation and map their competition continuously across the soft to hard X-ray range, rather than at a single edge.
The molecule is planar, and we have it lying in the $xy$ plane with $C_{2v}$ symmetry.
The field propagates in-plane along $\hat{k} = (0,1,0)$ with polarization $\hat{\varepsilon} = (1,0,0)$; at the sulfur K-edge the phase parameter $|\kgam|\,r_\mathrm{max} \approx 4.8$ places the interaction well beyond the dipole regime.
All runs use the PW92 LDA functional~\cite{perdewWang1992} in a non-relativistic treatment, a perturbative amplitude $A_0 = 10^{-4}$\,a.u.\ (doubling $A_0$ doubles every channel amplitude), and a Gaussian pulse of $\sigma_t = 0.020$\,fs ($78$\,eV spectral FWHM), with a velocity-gauge Crank--Nicolson propogation.
We scan fourteen carriers from $100$\,eV to $5$\,keV, bracketing the sulfur L$_{2,3}$, L$_1$, and K edges and the off-resonant gaps, with carriers at the carbon K-edge ($285$\,eV) and at the LDA-shifted sulfur $1s$ resonance ($2395$\,eV, see below).

We report each channel in electron units, $\chi_\mathrm{ch}(\om_0) = |\tilde{J}_\mathrm{ch}(\G_1,\om_0)|/|\tilde{A}(\om_0)|$, which divides out the pulse and expresses the response as a driven susceptibility (extraction and validation in the SI).
In these units the diamagnetic channel is the exact Thomson term of Eq.~\eqref{eq:JdiaG1}, $\chi_\mathrm{dia} = N_e = 44$, and the propagations reproduce this to better than $0.25\%$ at every carrier, an analytic check on the whole chain.
Because a bound core excitation has no decay path in the propagation, its bare spectral peak grows without bound with the trace length; we restore the physical linewidth by damping both channels at the natural core-hole width~\cite{krauseOliver1979} before the transform and propagate each carrier until the damped peak converges (details in SI).
The resonant $\chi_\mathrm{para}$ reported below are these lifetime-limited, converged values.

\begin{figure}[htbp]
  \centering
  \includegraphics[width=0.8\textwidth]{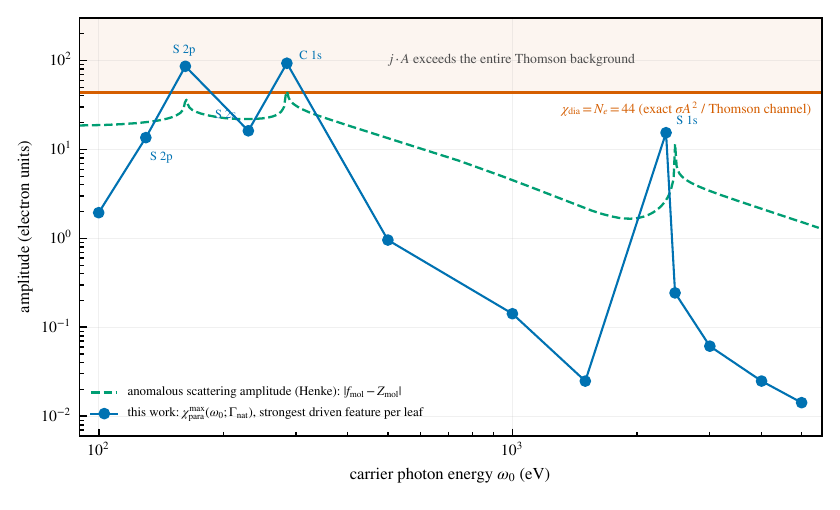}
  \caption{Diamagnetic versus paramagnetic response of thiophene, in electron units, for in-plane propagation ($\hat{k}\,\|\,\hat{y}$, $\hat{\varepsilon}\,\|\,\hat{x}$) at $A_0 = 10^{-4}$\,a.u.
  Blue: the strongest paramagnetic feature driven within the pulse bandwidth at each carrier, $\chi^\mathrm{max}_\mathrm{para}$, labeled by the core shell.
  Vermillion: the diamagnetic channel, exactly $\chi_\mathrm{dia} = N_e = 44$ (the Thomson background).
  Green dashed: the anomalous scattering amplitude $|f_\mathrm{mol}-Z_\mathrm{mol}|$ from the Henke tables~\cite{henke1993}, in the same units.
  The paramagnetic response crosses above the entire diamagnetic channel at the S\,$L_{2,3}$ and C\,K white lines and spikes at the sulfur $1s$ line without crossing; off resonance it decays parallel to the Henke curve.}
  \label{fig:thiophene}
\end{figure}

Figure~\ref{fig:thiophene} shows the outcome of these propagations with a few key takeaways.
First, the diamagnetic channel is a frequency-independent Thomson background at $N_e = 44$ as $\sigma\A^2$ channel contributes a flat floor since $|\tilde{A}(\om_0)|$ is fixed by the identical pulses and $N_e$ conserved.
Second, when off-resonance $\chi_\mathrm{para}$ decays with energy in step with the tabulated anomalous amplitude $|f_\mathrm{mol}-Z_\mathrm{mol}|$, staying far below the Thomson line; this numerically justifies the typical approach of neglecting the $\jj^\mathrm{para}\cdot\A$ term when calculating scattering in this off-resonant regime. 
Third, at the strong, narrow white lines of the S\,$2p$ and C\,$1s$ shells the lifetime-limited paramagnetic response ($86$ and $93$ electron units, Table~\ref{tab:sulfur}) exceeds the full 44-electron background, so the current term dominates the coherent response outright.
The $130$\,eV carrier, whose bandwidth reaches the L$_{2,3}$ lines only at its tails, sits between the regimes ($9.6$ units), this is comparable behavior that \citet{Rouxel2016} assign as near-resonant detunings.
The sulfur $1s$ line gives a paramagnetic spike ($15.4$ units) that does not cross the background, but it will contribute as a slight peak in the summed signal as with the Henke curves~\cite{henke1993}.
The S K shell holds only two electrons and its width is an order of magnitude larger than that of the S L$_{2,3}$ shell, so lifetime-limited dominance is out of reach even on resonance; just above the K threshold ($2472$\,eV) the bound-line strength is gone.
Normalizing each line to its own shell's charge-density weight gives channel ratios between $8$ and $14$: the current term dominates every resonant pathway which is our channel-resolved form of the result from \citet{Rouxel2016} (additional details in the SI).
The absorptive response is carried entirely by the paramagnetic channel since $\chi_\mathrm{dia}$ is real.

\begin{table}[htbp]
  \centering
  \begin{tabular}{lrrrr}
    \toprule
    resonance & $\om_0$ (eV) & $\chi_\mathrm{para}$ ($e^-$) &
      $R_\mathrm{line}=N_e/\chi_\mathrm{para}$ & shell ratio $P$ \\
    \midrule
    S\,L$_{2,3}$ ($2p$)      & 162  & $86\pm3$ & $0.51$ & $14$ \\
    S\,L$_1$ ($2s$)          & 230  & $16$     & $2.7$  & $8$  \\
    C\,K ($1s$)              & 285  & $93$     & $0.47$ & $12$ \\
    S\,K ($1s$, LDA)         & 2395 & $15.4$   & $2.9$  & $8$  \\
    \bottomrule
  \end{tabular}
  \caption{Lifetime-limited, trace-length-converged paramagnetic line strengths in electron units at specific carrier energies $\om_0$ and its diamagnetic-to-paramagnetic line ratio $R_\mathrm{line} = N_e/\chi_\mathrm{para}$ (when $<1$ the current term dominates the total coherent response) as well as the channel ratio $P = \chi_\mathrm{para}/n_\mathrm{shell}$ that weighs the response against the resonant shell's charge-density.}
  \label{tab:sulfur}
\end{table}

These results reproduce the two robust elements of the \citet{Rouxel2016} picture, the frequency-independent $\sigma\A^2$ background and the resonant dominance of $\jj^\mathrm{para}\cdot\A$, non-perturbatively and as a continuous function of photon energy.
Two features of the setup shape the comparison.
First, the PW92 sulfur $1s$ eigenvalue places the LDA K-edge resonance near $2395$\,eV, about $80$\,eV below the experimental edge~\cite{henke1993,perdewZunger1981}; the resonant carrier is placed there, and in this in-plane polarization the strongest K-region line lies at $2351$\,eV while the out-of-plane $1s\to\pi^*$ transition appears only weakly.
Second, off-resonance our $\chi_\mathrm{para}$ runs a factor of $5$ to $90$ below the Henke curve: an FHI-aims `tier-1' basis exhausts only a third of the velocity-gauge sum rule (SI), so the off-resonant amplitudes are lower bounds, while the resonant bound-line strengths, carried by states the basis describes, are the robust content.
A scalar-relativistic treatment (atomic ZORA~\cite{aims}) at the L edges shifts the L$_{2,3}$ line strength by ${\sim}18\%$ relative to the non-relativistic baseline, without altering any conclusion; the $1.2$\,eV sulfur $2p$ spin-orbit splitting is unresolved at the $78$\,eV probe bandwidth, and the second-variational spin-orbit correction available in FHI-aims is post-SCF and does not enter the real-time propagation here.
The Rouxel crossover, a flat diamagnetic background overtaken by a resonant paramagnetic channel, is reproduced in full, with the total-current crossover realized at the S\,L$_{2,3}$ and C\,K white lines.

\FloatBarrier
\subsection{Diamond second-harmonic generation at 1\,keV: a symmetry-forbidden forward channel and its intensity scaling}
\label{sec:diamond}

Here we extend our methodology to nonlinear X-ray techniques, studying second-harmonic generation in diamond at hard X-ray energies. 
In this technique, the signal is generated at twice the input frequency, with the signal proportional to the square of the field strength (second-order process).  
We use the reciprocal-space-resolved current diagnostic and the photon-momentum sideband coupling in a periodic crystal at X-ray energies, where the diamagnetic signal is expected to dominate per the mechanism of \citet{Rouxel2016} and the cold-plasma picture of Freund and Eisenberger~\cite{freund1970,eisenberger1971}.
The system used here is diamond, the same medium in which \citet{shwartz2014} observed the first hard-X-ray SHG signal at $7.3$\,keV, and the high symmetry of the structure has a variety of benefits for simplifying analysis.~\cite{tom1983} 
We choose $\om = 1000$\,eV, well above the carbon K-edge at $284$\,eV~\cite{henke1993} (placing the calculation deep in the diamagnetic regime) and computationally tractable while still firmly beyond the dipole limit.
The surface sensitivity of X-ray second-harmonic generation and its dependence on photon energy were recently characterized by \citet{schacher2026} using a combination of (ED) RT-TDDFT and analytic methods, and by $1000$\,eV it is predicted to be bulk-dominated. 
The real-time, beyond-dipole calculation presented here provides a deeper analysis, separating the symmetry-forbidden forward channel and the competing surface and bulk contributions directly from the propagated current.

\begin{figure}[htbp]
  \centering
  \includegraphics[width=0.75\textwidth]{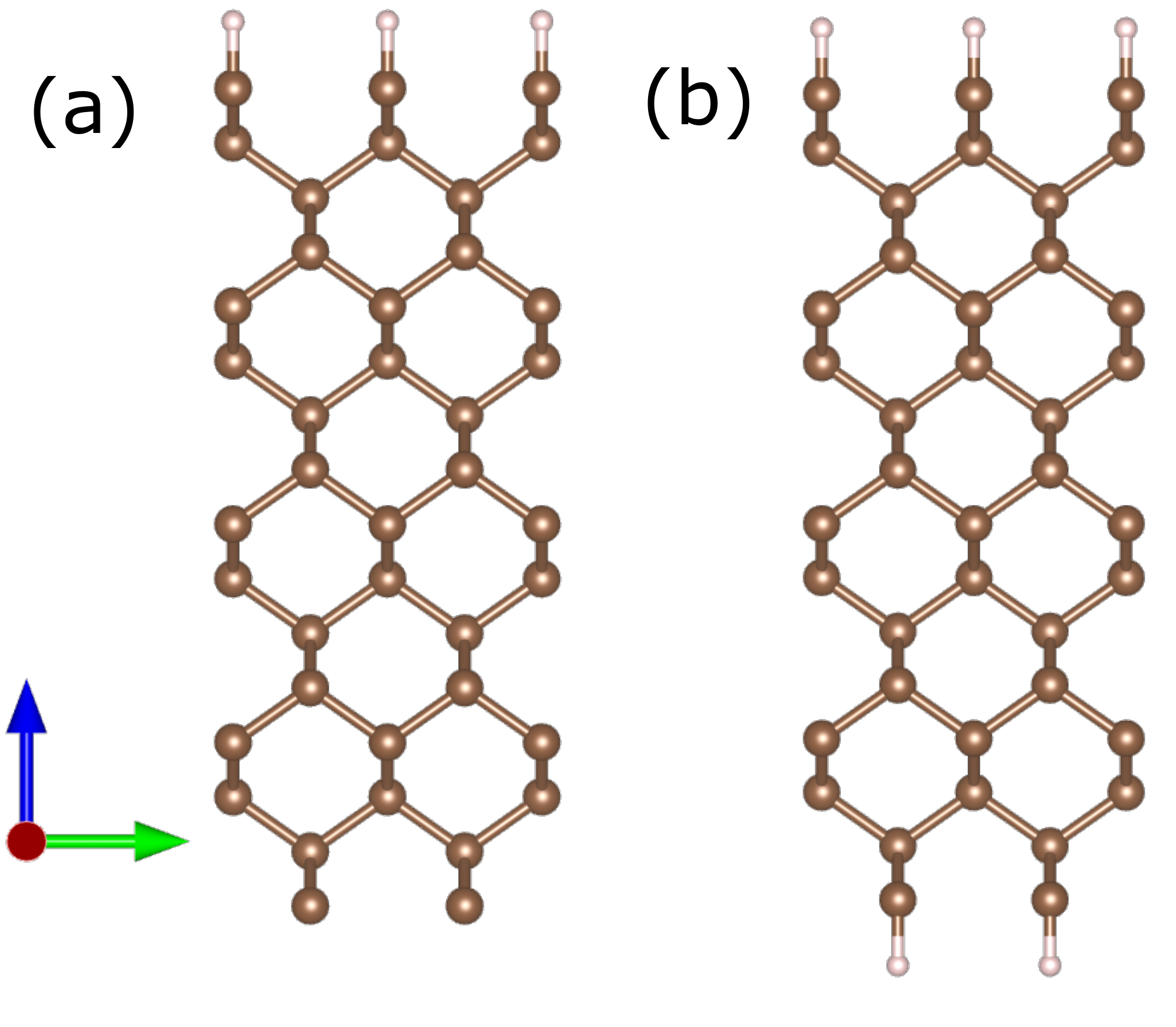}
  \caption{
  Diamond(001) slabs for the 1\,keV SHG benchmark (16 C layers; C brown, H pink).
  \textbf{(a)}~Asymmetric (one face H-terminated, one bare): inversion broken, forward SHG dipole-allowed. 
  \textbf{(b)}~Symmetric (both faces H-terminated): centrosymmetric, forward SHG dipole-forbidden.
  }
  \label{fig:diamond_structure}
\end{figure}

The calculations use three geometries that share an identical $xy$ surface (the diamond $(001)$ $1\times1$ cell, $a/\sqrt{2} = 2.523$\,{\AA}) and a common Gaussian pulse: a fully periodic bulk cell (four C atoms spanning one cubic period along $[001]$, $a = 3.567$\,{\AA}, $12\times 12\times 12$ k-mesh); a $(001)$ symmetric slab of sixteen carbon layers with hydrogen passivating each face ($12 \times 12 \times 1$ k-mesh, $50$\,{\AA} $\vec{c}$ with $\sim 17$\,{\AA} vacuum on each side); and an otherwise identical asymmetric slab with the hydrogen removed from one face to lower the symmetry of the system (Fig.~\ref{fig:diamond_structure}).
The hydrogen atoms are added to passivate the carbon atoms and remove dangling broken bonds.
The observable in this case is the macroscopic forward second harmonic, the $\G = 0$ Fourier component $|J_z(\G{=}0,2\om_0)|$ of the propagated current projected on the polarization.
For the centrosymmetric bulk and symmetric slab, this forward channel is strictly forbidden by inversion symmetry in the dipole limit, so any coherent $2\om_0$ amplitude it carries under full minimal coupling is pure beyond-dipole physics.
The asymmetric slab breaks inversion and adds an electric-dipole surface $\chi^{(2)}$ on top of that bulk response; the surface contribution is isolated as the difference between the asymmetric and symmetric forward signals, the common centrosymmetric component canceling because the two slabs share identical carbon positions.
The field propagates in-plane along $\hat{k} = (1,0,0)$ with polarization $\hat{\varepsilon} = (0,0,1)$ along the surface normal, giving $\G_1 = (\om/c)\hat{x}$ with $|\kgam| = 0.506$\,{\AA}$^{-1}$ and $\G_2 = 2\G_1$.

The pump amplitude is swept across five values log-spaced from $A_0 = 1\times 10^{-5}$ to $1\times 10^{-3}$\,a.u., spanning two orders of magnitude, with the PBE functional~\cite{perdewBurkeErnzerhof1996}, velocity-gauge Crank--Nicolson propagation, $\sigma_t = 0.200$\,fs, $T = 1.000$\,fs, and the photon-momentum sideband coupling common to all runs.
Each geometry is also propagated in a dipole-limit reference mode with the spatial phase $e^{i\kgam\cdot\rr}$ and the photon-momentum sideband coupling switched off; for the symmetric slab this reference is taken at every amplitude, so the residual (numerically symmetry-broken) forward signal can be bounded across the full range, and single references at $A_0 = 10^{-4}$\,a.u.\ are run for the bulk and asymmetric geometries.

\begin{figure}[htbp]
  \centering
  \includegraphics[width=\textwidth]{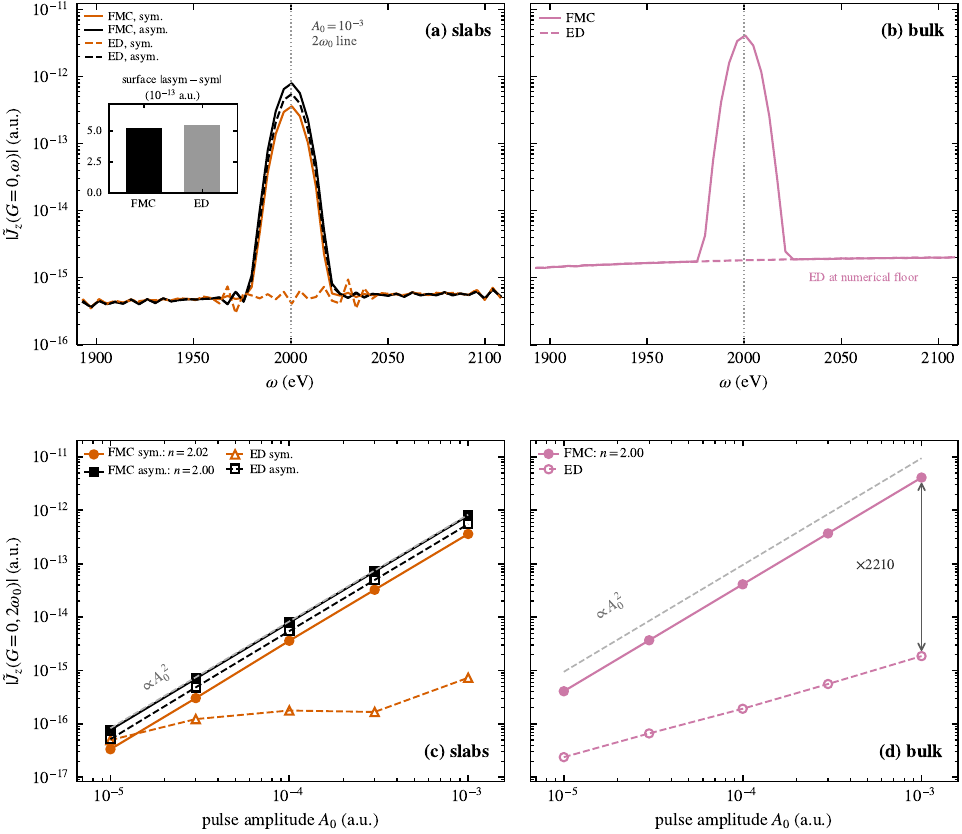}
  \caption{Forward ($\G = 0$) second-harmonic generation of diamond at $\om_0 = 1000$\,eV, the channel forbidden by inversion symmetry in the dipole approximation, across the $(001)$ slabs with hydrogen passivation on both faces (symmetric, H/H) or just one face (asymmetric, H/bare) and the bulk.
  Color denotes geometry; solid lines and filled markers are full minimal coupling (FMC), dashed lines and open markers are the dipole-limit (ED) reference.
  Left column, the two slabs; right column, the bulk.
  \textbf{(a,\,b)} Spectral amplitude of the macroscopic current near the second harmonic at $A_0 = 10^{-3}$\,a.u.
  In the slabs (a) the symmetric-slab FMC line sits a factor of $\sim 490$ above its ED floor, and removing one hydrogen face (asymmetric) lifts the ED reference to a dipole-allowed surface $\chi^{(2)}$ whose magnitude $|\mathrm{asym}-\mathrm{sym}|$ at the peak is equal in FMC and ED to within $5\%$ (inset), confirming it is dipole-allowed; in the bulk (b) the ED reference carries no second-harmonic peak and lies at the numerical floor.
  \textbf{(c,\,d)} Amplitude of the $2\om_0$ line versus pump amplitude over two decades.
    The FMC series and the dipole-allowed asymmetric-slab ED series scale quadratically (fitted exponents $n$).
    At the symmetry-forbidden channels the FMC-to-ED gap is the beyond-dipole enhancement, $\sim 490$ for the symmetric slab (c) and $\sim 2200$ for the bulk (d, no surfaces and extra glide symmetries), while the asymmetric slab's FMC signal sits a factor of $\sim 1.4$ above its dipole-allowed surface-$\chi^{(2)}$ reference.}
  \label{fig:diamond}
\end{figure}

In a centrosymmetric crystal, the forward ($\G = 0$) second harmonic vanishes in the dipole limit; under full minimal coupling the symmetric slab instead develops a coherent line at $2\om_0 = 2000$\,eV (Fig.~\ref{fig:diamond}a), a factor of $\sim 490$ in amplitude and $\sim 2.4\times10^{5}$ in intensity, above the dipole-limit reference of the identical geometry.
The line scales as $A_0^2$ over the full two order of magnitude amplitude range (Fig.~\ref{fig:diamond}c, fitted exponent $n = 2.02$), the signature of a coherent second-order process. 
Producing this forward channel at all is the central demonstration of the method: it is symmetry-forbidden to standard dipole-approximation RT-TDDFT, and it appears here only because the spatial structure of the field is retained.

The dipole-limit reference provides a null result within numerical error.
Any residual is set by inversion breaking in the self-consistent field and the integration grid, sits roughly three orders of magnitude below the FMC line, and across the amplitude range it scales sub-quadratically (exponent $\sim 0.9$ for the bulk, $\sim 0.5$ for the symmetric slab) rather than as the coherent $A_0^2$ of a true second-order process.
It therefore bounds, rather than measures, the forbidden channel, and we quote the FMC-to-reference suppression factor rather than an absolute zero.
Two independent checks establish that the FMC line is physical and not a numerical residue.
First, the fully periodic bulk crystal, which carries additional symmetries and no surface states, gives a cleaner separation still. The bulk FMC $\G = 0$ second harmonic scales as $A_0^{2.02}$ and exceeds the dipole-limit reference by a factor of $\sim 2200$ (Fig.~\ref{fig:diamond}d), against $\sim 490$ for the symmetric slab and $\sim 1.4$ for the asymmetric slab (discussed below) whose dipole-limit reference already carries the surface $\chi^{(2)}$ (Fig.~\ref{fig:diamond}c).
Second, a standard ED approximation RT-TDDFT propagation reproduces the macroscopic linear current of the dipole-limit reference to $\sim 1\%$ across the spectrum and leaves the forward second harmonic at the same numerical floor, confirming both that the implementation reduces correctly to standard RT-TDDFT in the dipole limit and that the forward SHG line appears only with the full spatial field.

\begin{figure}[htbp]
  \centering
  \includegraphics[width=0.55\textwidth]{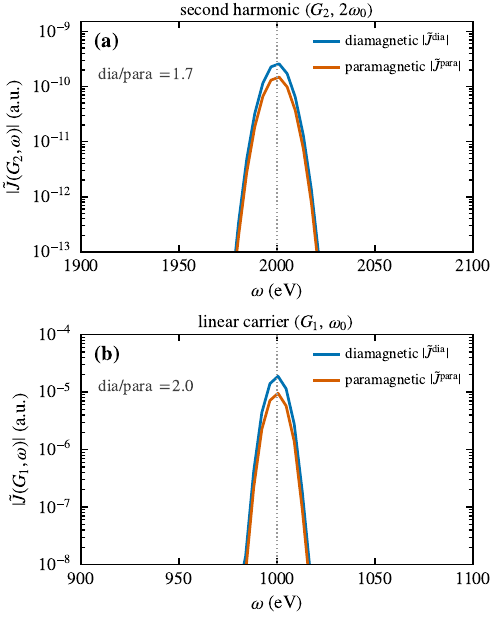}
  \caption{G-resolved channel decomposition of bulk diamond at $\om_0 = 1000$\,eV ($A_0 = 10^{-3}$), a mechanistic diagnostic.
  The current at the photon wavevector splits into a paramagnetic (anomalous) and a diamagnetic (Thomson) part.
  \textbf{(a)} At the second harmonic ($\G_2$, $2\om_0$), the diamagnetic channel exceeds the paramagnetic by a factor of $\sim 1.7$.
  \textbf{(b)} At the linear carrier ($\G_1$, $\om_0$), the diamagnetic response is larger by a factor of $\sim 2.0$. The hard X-ray cold-plasma Thomson response is where the diamagnetic character comes from, opposite to the resonant paramagnetic dominance at the thiophene sulfur K-edge (Fig.~\ref{fig:thiophene}).
  These windowed-$\G$ channels are single-cell projections, a mechanism diagnostic and not the macroscopic SHG observable of Fig.~\ref{fig:diamond}.}
  \label{fig:diamond_channels}
\end{figure}

Breaking inversion symmetry by removing the hydrogen passivation of one face makes forward SHG dipole-allowed through a surface $\chi^{(2)}$, seen directly as the asymmetric slab's dipole-limit reference (Fig.~\ref{fig:diamond}a, black dashed) rising a factor of $750$ above the symmetric ED floor and, unlike that floor, scaling as $A_0^{2.01}$, the coherent signature of a genuine surface $\chi^{(2)}$.
The surface contribution separates cleanly: the asymmetric-minus-symmetric difference signal is the same under full minimal coupling and in the dipole limit to within $5\%$ ($5.3$ versus $5.5\times10^{-13}$\,a.u.), so the added surface $\chi^{(2)}$ is a dipole-limit object riding on top of the beyond-dipole response, and the surface and beyond-dipole contributions are separately recoverable.
Because the surface term is comparable to the bulk at $1$\,keV, the forward SHG of diamond is already bulk- and plasma-dominated rather than surface-dominated at this energy.
Consistent with the cold-plasma expectation~\cite{freund1970,eisenberger1971,Rouxel2016}, the G-resolved channel decomposition (Fig.~\ref{fig:diamond_channels}) is diamagnetic-dominated: the Thomson channel exceeds the paramagnetic by factors of $\sim 2.0$ at the linear carrier and $\sim 1.7$ at the second harmonic, in contrast to the resonant paramagnetic dominance at the thiophene core edge.
The FMC forward-SHG magnitude is moreover robust to the surface termination in both the symmetric and asymmetric cases ($\sim 15\%$ between the two centrosymmetric terminations and $\sim 11 \%$ between the two inversion-broken ones), so the beyond-dipole signal is a bulk property rather than a surface one.
The dipole-limit (ED) floors of the two centrosymmetric terminations are equally clean, with suppression factors of $\sim 490$ (H/H) and $\sim 520$ (He/He), so the forbidden-channel null is robust to whether the faces are covalently passivated or physisorbed, despite the dangling-bond surface states the helium leaves intact (Figure~S2 and Table~S1, Supporting Information).

In a recent perspective, \citet{shen2025} questions reported demonstrations of soft X-ray second-harmonic generation as a surface-specific probe on symmetry and phase-matching grounds.
Specifically, that the transmitted (forward) geometry of those experiments lets the electric-quadrupole bulk second-order susceptibility dominate the dipole-allowed surface term.
Thus, the signal is bulk-dominated rather than genuinely surface-specific, and interpreting it requires a theory that includes the bulk contribution.
Our forward ($\G = 0$) second-harmonic calculation is exactly such a bulk-inclusive decomposition from first principles, and it agrees with Shen's argument in this far from resonant case and corroborates our previous analysis of the energy dependence of X-ray SHG surface sensitivity~\cite{schacher2026}.
As this method offers a direct, bulk-inclusive route to the surface-to-bulk ratio that any claim of nonlinear X-ray surface specificity must establish.
Future work is planned to investigate this further in other systems with multiple resonances against photon energy and geometry to identify the conditions under which these signals are genuinely surface-specific.

\FloatBarrier
\subsection{Carbon monoxide oxidation on Ru(0001) at the oxygen K-edge: full minimal coupling on a metallic catalytic surface}
\label{sec:ostrom}

The last example studies another surface-chemistry problem: the oxygen K-edge X-ray absorption of CO and atomic O coadsorbed on Ru(0001) along the CO oxidation pathway following the experimental work of \citet{ostrom2015} and and \citet{Murano2026}.
It is a different test of the formalism than any of the preceding benchmarks, being the only one on a conducting d-band substrate, where the core-level response sits on top of a metallic continuum.
The beyond-dipole regime is relatively weak but unambiguous across the near-edge region: at the O K-edge $\om \approx 525$\,eV, $|\kgam| = 0.27$\,{\AA}$^{-1}$, so the spatial phase advances by $|\kgam|\,R_\mathrm{ads} \approx 0.7$ over the $\sim 2.6$\,{\AA} CO+O complex and by $|\kgam|\,d_\mathrm{slab} \approx 1.7$ across the four-layer Ru slab, large enough that the dipole approximation cannot be assumed. In a variety of previous cases, quadrupole transitions have been seen at the O K-edge.~\cite{FDG}

\begin{figure}[htbp]
  \centering
  \includegraphics[width=0.75\textwidth]{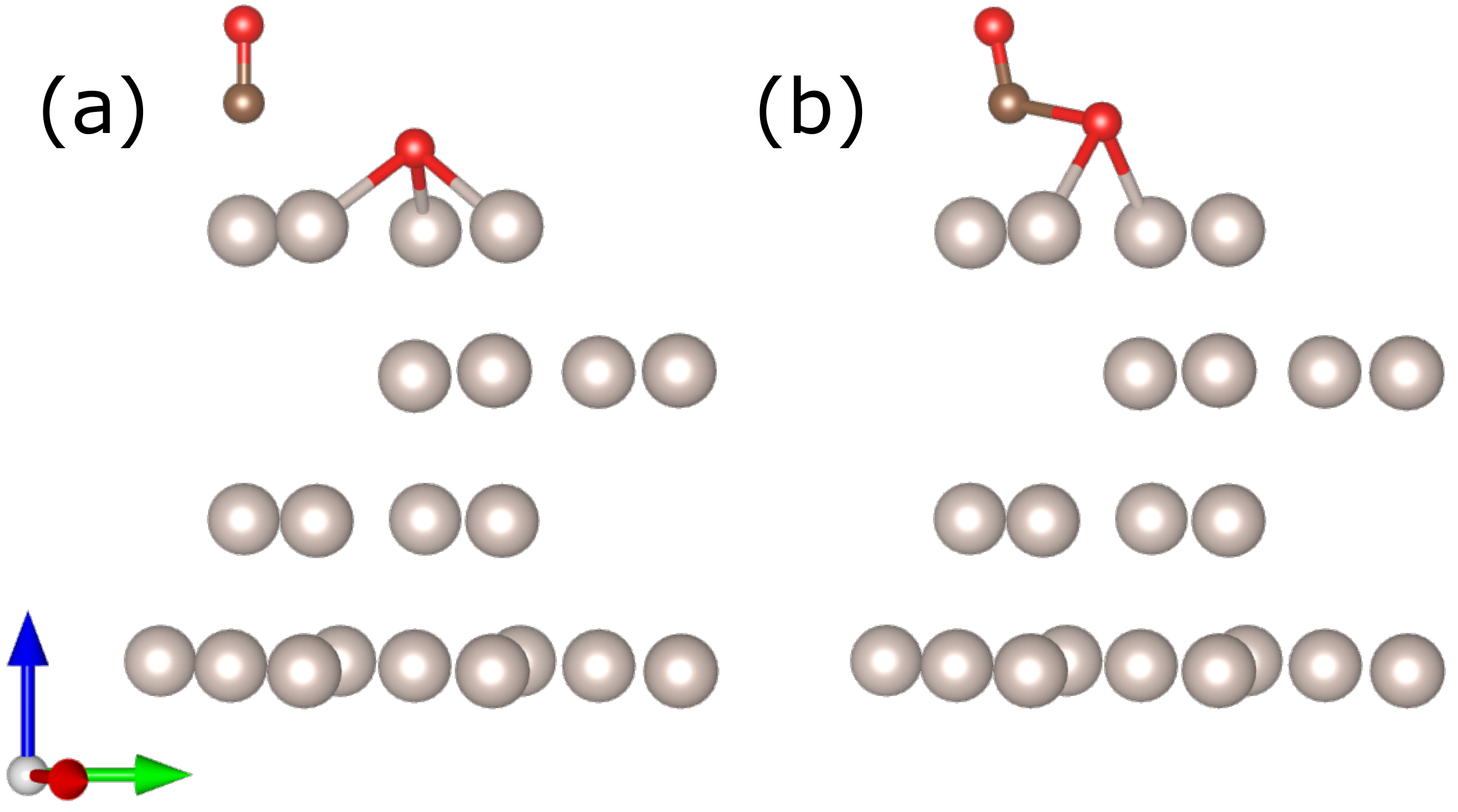}
  \caption{CO oxidation benchmark on Ru(0001) 4--layer $2\times2$ Ru slab (Ru taupe, C brown, O red).
  \textbf{(a)} Initial state (IS) structure with upright CO and atomic O, separated by $\approx3.2$\,\AA. 
  \textbf{(b)} Transition state (TS1) with atomic O bonded to the CO carbon (C--O $=1.72$\,\AA{} and O--C--O $\approx 114^{\circ}$), en route to CO$_2$\cite{ostrom2015}.}
  \label{fig:ostrom_struct}
\end{figure}

The system is a Ru(0001) $2\times 2$ four-layer slab (16 Ru) with CO and atomic O adsorbed on the top face only, in the initial-state (IS) geometry of \"Ostr\"om et al.\: CO at the atop site and atomic O at the hcp hollow above a second-layer Ru atom (Fig. ~\ref{fig:ostrom_struct}.
The field propagates laterally, $\hat{k} = (0,1,0)$, with surface-normal polarization $\hat{\varepsilon} = (0,0,1)$, the $p$-polarized surface-XAS geometry, so the spatial phase $e^{i(\om/c)y}$ varies across the adsorbate complex and the polarization probes the O\,$2p^*$ and broader $\sigma^*$ resonances perpendicular to the surface.
Because the slab is periodic in-plane it carries no macroscopic dipole moment, so the absorption lineshape cannot be read from a dipole spectrum as in the molecular benchmarks above.
We instead reconstruct it pointwise, from a scan of $17$ narrowband carriers $\om_0$ spanning $509.5$ to $546$\,eV, each a Gaussian pulse of temporal width $\sigma_t = 0.2$\,fs (spectral FWHM $\sim 7.4$\,eV, narrow enough for carrier selectivity), velocity-gauge Crank--Nicolson propagation at $\Delta t = 4\times 10^{-4}$\,fs, perturbative amplitude $A_0 = 10^{-4}$\,a.u., and the PBE functional~\cite{perdewBurkeErnzerhof1996}.
Each carrier is run as a dipole-approximation reference (ED) and a full minimal coupling (FMC) pair.

In this case, two single-number observables are extracted at each carrier that are sensitive to complementary parts of the response.
The first is the absorbed energy $\dEabs(\om_0)$, the drift in the total electronic energy across the pulse, which is proportional to the absorptive part $\mathrm{Re}\{\sigma(\om_0)\}$ of the conductivity.
The second is the peak driven current $|\tilde{J}_z(\G{=}0,\om_0)|$, the polarization-projected macroscopic current at the carrier, which is the standard current-based XAS observable; it is dominated by the reactive (Thomson) part of the response and so tracks the dipole-allowed near-edge lineshape.
Plotting either against $\om_0$ across the scan traces the lineshape, and the FMC$-$ED difference at each carrier isolates the beyond-dipole correction as a function of photon energy.

\begin{figure}[htbp]
  \centering
  \includegraphics[width=0.6\textwidth]{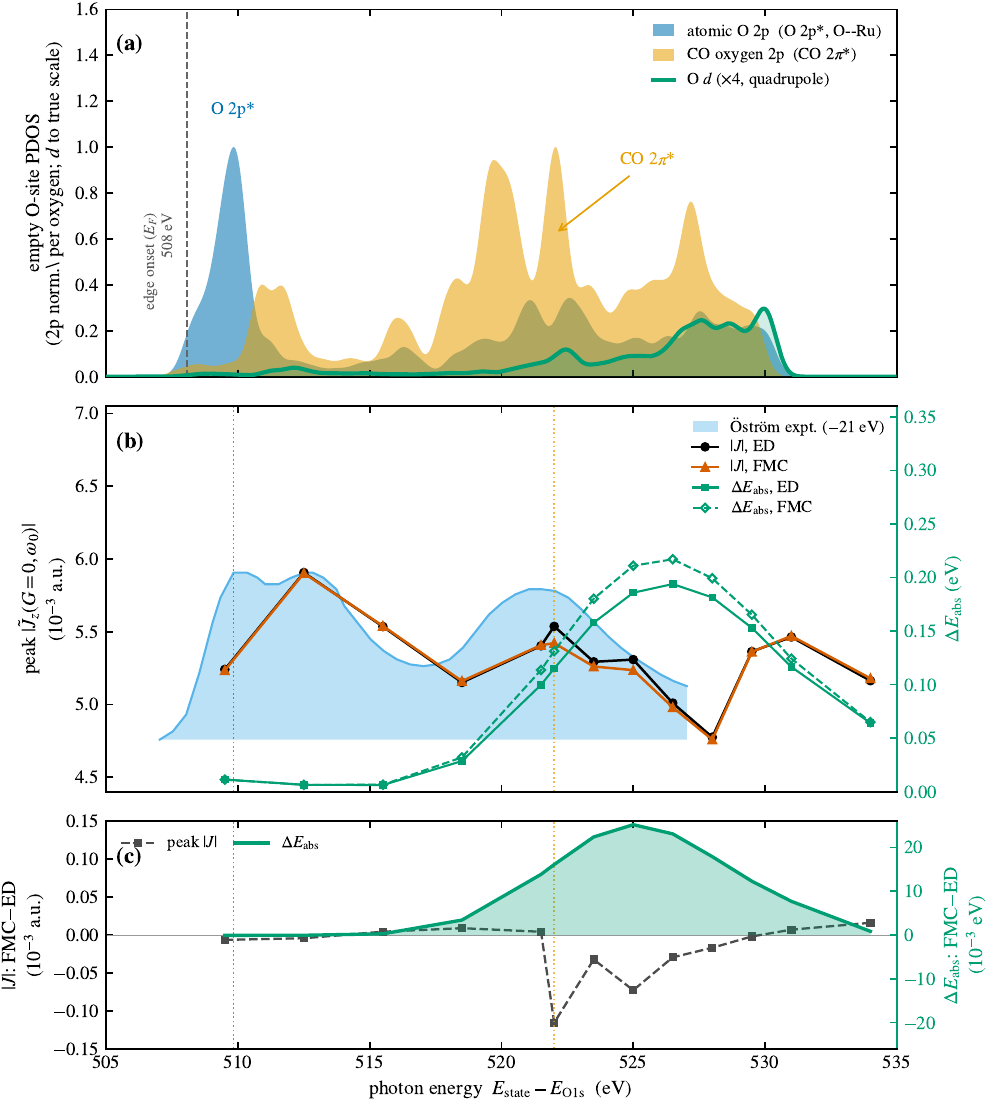}
  \caption{Oxygen K-edge of CO and atomic O coadsorbed on Ru(0001) (initial state): full minimal coupling RT-TDDFT against the static (unpumped) experimental spectrum of \citet{ostrom2015}.
  \textbf{(a)} Empty O-site projected density of states (ground-state PBE).
  The $2p$ (dipole) channel, resolved per oxygen, places the chemisorbed atomic O\,$2p^*$ (O--Ru antibonding) at the $508$\,eV Fermi-level edge onset and the CO\,$2\pi^*$ about $12$\,eV higher; the O\,$d$ projection (green, scaled $\times 4$; $d/p \approx 0.25$ at the peak) is the quadrupole channel ($\mathrm{O}\,1s \to d$, $\Delta\ell = 2$) opened only beyond the dipole approximation, negligible at the O\,$2p^*$ threshold and peaking at $522$--$530$\,eV.
  \textbf{(b)} Lineshapes of the two observables, with the consistent left\,=\,current, right\,=\,energy axis convention.
  Left axis: the peak driven current $|\tilde{J}_z(\G{=}0,\om_0)|$ (ED and FMC, which overlap), reproducing the experimental two-feature near-edge structure (O\,$2p^*$ and CO\,$2\pi^*$).
  Right axis (green): the absorbed energy $\dEabs$ (ED solid, FMC dashed), which instead peaks higher, at the metallic white line ($\sim 526$\,eV), above the adsorbate features.
  The experiment (blue) is rigid-shifted by $-21$\,eV to align the O\,$2p^*$ leading feature.
  \textbf{(c)} Beyond-dipole correction (FMC$-$ED) of each observable, on matching axes.
  Left (grey): the current difference, negligible (within $\sim 2\%$) because the current amplitude is dominated by the dipole-allowed response.
  Right (green): the absorbed-energy difference, near zero at the dipole-only O\,$2p^*$ threshold and rising to $+25\times10^{-3}$\,eV ($+14\%$) across the white line, coinciding with the O\,$d$ quadrupole channel of panel (a).
  The beyond-dipole effect is thus purely absorptive.}
  \label{fig:ostrom}
\end{figure}

Absolute energies are placed on the PBE photon scale $E_\mathrm{state} - E_\mathrm{O\,1s}$ built from the computed Kohn--Sham eigenvalues, on which the leading O\,$2p^*$ feature falls at $509.8$\,eV.
This is $\sim 21$\,eV below the experimental O\,$2p^*$ position of $530.8$\,eV, an expected consequence of DFT underestimating the band gap~\cite{perdewZunger1981}.
We therefore calibrate to the computed eigenvalue and rigid-shift the experimental spectrum by $-21$\,eV to align the leading feature, rather than referencing the experimental edge. 
Figure~\ref{fig:ostrom}A shows the projected density of states.
PBE additionally over-separates the O\,$2p^*$ and CO\,$2\pi^*$ resonances ($\sim 12$\,eV here against the experimental $\sim 3$\,eV), consistent with the known tendency of semilocal functionals to misplace the CO frontier ($2\pi^*$/$5\sigma$) levels relative to the metal d-band~\cite{feibelman2001}; part of this discrepancy also reflects the relative O\,$1s$ core-level shift between the two inequivalent oxygen atoms.

The driven current (Fig.~\ref{fig:ostrom}b, left axis) reproduces the experimental near-edge lineshape: a leading O\,$2p^*$ peak near $512$\,eV and a CO\,$2\pi^*$ shoulder near $521$\,eV, matching the two features of the shifted \"Ostr\"om spectrum.
Here, the ED and FMC current curves lie on top of one another, and their difference (Fig.~\ref{fig:ostrom}c, left axis) stays within $\sim 2\%$ across the whole scan with a maxima at the CO\,$2\pi^*$ feature.
In the reactive channel that carries the lineshape, the beyond-dipole correction is small, well below the absorptive correction, as that amplitude is set by the dipole-allowed response.
The absorbed energy (Fig.~\ref{fig:ostrom}b, right axis) tells a different story; it peaks not at the adsorbate resonances but at $526.5$\,eV, where the absorptive $\mathrm{Re}\{\sigma\}$ is largest.

The beyond-dipole physics is most important in this absorptive channel.
The FMC$-$ED difference of the absorbed energy (Fig.~\ref{fig:ostrom}c, right axis) is essentially zero at the O\,$2p^*$ threshold ($509$--$513$\,eV), where the only available transition is the dipole-allowed $\mathrm{O}\,1s \to 2p$, and turns on across the higher energy region, reaching +14\% of the ED value at $525$\,eV.
That onset coincides exactly with the empty O\,$d$ projected density of states of Fig.~\ref{fig:ostrom}a and is the signature of the quadrupole-allowed $\mathrm{O}\,1s \to d$ ($\Delta\ell = 2$) transition, which is forbidden in the dipole limit and opened only by the retained spatial phase of the field.
These $d$ states are not an isolated atomic resonance on O: the empty Ru\,$4d$ band begins directly at the Fermi level ($508$\,eV on the photon scale, the edge onset of Fig.~\ref{fig:ostrom}a) and dominates the total empty density of states across the entire scan window by roughly an order of magnitude, and the O\,$d$ curve of Fig.~\ref{fig:ostrom}a is the small atom-local projection of this band onto the absorbing oxygen ($d/p \approx 0.25$ at its $522$--$530$\,eV maximum).
Given the atomistic localization of the core-level hole, the O-site $d$ character predominantly carries O K-edge quadrupole intensity; the Ru\,$4d$ band acts as the hybridization partner that supplies it, the same O--Ru mixing that binds the adsorbates to the surface.
The effect is purely absorptive, present in $\dEabs$ but not appreciable in the current amplitude, consistent with a channel that adds oscillator strength to the absorption without altering the dominant reactive response.

The same purely-absorptive signature recurs when the calculation is repeated at the transition-state geometry (TS1) as well as for the individual, isolated adsorbates on the slab in their IS geometry.
In TS1, the two adsorbates have migrated towards one another to form the roughly surface-parallel O--C bond of the oxidation step ($1.72$\,{\AA}, down from $3.2$\,{\AA} in the IS), with the atomic O climbing out of its hollow site and the CO axis tilting $\sim 14^{\circ}$ off the surface normal on the way to CO$_2$ formation (Fig.~\ref{fig:ostrom_struct}b).
In these cases, the beyond-dipole correction grows from $\sim 11\%$ (TS1) to $\sim 19$--$21\%$ (lone O and lone CO); these confirm that the $s\to d$ quadrupole enhancement is a generic feature of the O K-edge O\,$d$ states mixing with the Ru rather than an artifact of the IS co-adsorbate geometry (Figs.~S3 and S4, Supporting Information).

This shows the importance of beyond-dipole physics on a conducting d-band substrate, with the two observables cleanly separating the dipole-allowed reactive current, where FMC and ED agree to within $\sim 2\%$, from the absorptive energy.
This split follows the multipole expansion laid out in the benzene benchmark (Sec.~\ref{sec:benzene}).
There, the even second-order term $\tfrac{1}{2}(\kgam\cdot\rr)^2$, in phase with the carrier, produced the weak percent-level correction to the driven dipole, while the odd first-order term contributed no in-phase dipole for the centrosymmetric ring.
Here the complementary term does the work: the reactive current again changes only weakly, but the odd first-order term $i\,\kgam\cdot\rr$, in quadrature with the carrier, opens the quadrupole O\,$1s\to d$ channel in the absorptive observable and adds $\sim 14\%$.
The two benchmarks thus exercise the two leading terms of the same expansion in complementary observables.

\FloatBarrier
\section{Conclusions}\label{sec:conclusion}

We have presented an all-electron, numeric-atom-centered-orbital implementation of full minimal coupling in real-time TDDFT in FHI-aims, retaining the spatial structure of the vector potential $\A(\rr,t)$ in both the paramagnetic $\A\cdot\hat{p}$ and diamagnetic $\A^2$ couplings, together with a reciprocal-space-resolved current diagnostic and a photon-momentum sideband coupling that recovers the density response at the photon wavevector in periodic systems.
Retaining the photon momentum gives first-principles access to processes invisible to the dipole limit.
The first-order $\kgam\cdot\rr$ term that the implementation propagates is what produces nondipole forward/backward asymmetries in photoionization and photoemission angular distributions~\cite{krassig1995}, which an all-electron treatment can compute at relevant edges across the periodic table.
The reciprocal-space-resolved current and the photon-momentum sideband coupling are, in turn, built for nonlinear processes that appear at finite reciprocal-space wavevector, including Bragg-geometry nonlinear diffraction and X-ray--optical wave mixing or parametric down-conversion~\cite{freund1970,eisenberger1971}, where the signal is carried at a reciprocal-lattice vector rather than in the forward direction.

There are a few key approximations to improve cost-effectiveness of the method.
First, the periodic photon-momentum sideband coupling propagates only a single pair of off-diagonal $\kk$-blocks ($n = \pm 1$) under a $\kk$-diagonal free-propagation operator and a first-order operator splitting, and keeps only the on-site ($R = 0$) term of the lattice sum.
This suffices for the leading beyond-dipole density response and the macroscopic forward ($\G = 0$) second harmonic, but higher sideband orders and the corresponding higher harmonics are not captured, and the windowed finite-$\G$ current channels are single-cell projections to be read as mechanism diagnostics rather than macroscopic observables.
Studies on topics such as high harmonic generation or a Bragg-geometry second-harmonic channel~\cite{shwartz2014}, will need to track higher sideband channels ($|n| > 1$) and go beyond the single-cell ($R = 0$) lattice sum to capture higher harmonics and turn the windowed finite-$\G$ currents into genuine macroscopic observables.
Second, the driving field is a prescribed beyond-dipole perturbation without the self-consistent back-action of the induced field that a coupled Maxwell--TDDFT treatment provides~\cite{bonafe2025}.
This will underestimate the correction when the induced field is strong, and should be incorporated when a quantitative response in the strong-field regime is needed.
Lastly, for symmetry-forbidden channels, the dipole-limit reference is a residual set by inversion breaking in the self-consistent field, integration grid, and finite system size rather than an exact null; thus we report suppression factors instead of absolute zeros as we anticipate these problems will persist even if not performed on a grid.

This technique as implemented here can be used to understand a variety of high energy physical processes in both the gas phase and condensed matter, as is demonstrated in four benchmarks spanning $100$\,eV to $5$\,keV that highlight that the retained spatial phase is physically consequential.
At the carbon K-edge of benzene, the beyond-dipole correction to the induced dipole agrees with the self-consistent Maxwell--TDDFT result~\cite{bonafe2025}: an RMS difference of $1$--$3.5\%$ of the dipole response across the scan, within the geometric $\tfrac{1}{2}(\kgam r)^2$ estimate. 
At the sulfur K-edge of thiophene, the G-resolved current reproduces the diamagnetic-versus-paramagnetic crossover~\cite{Rouxel2016} non-perturbatively and resolves it continuously in photon energy: the diamagnetic Thomson amplitude is flat to $2\%$ from $100$\,eV to $5$\,keV while the paramagnetic channel is resonantly enhanced at the core edges, while the paramagnetic channel is resonantly enhanced at the core edges and overtakes the Thomson background at the sulfur $2p$ and carbon $1s$ edges.\cite{freund1970}.
Full minimal coupling produces the forward ($\G = 0$) second harmonic of diamond at $1$\,keV that is strictly forbidden to dipole-limit RT-TDDFT in a centrosymmetric crystal: a coherent $2\om_0$ line scaling as $A_0^2$ and standing a factor $\sim 8$ (symmetric slab) to $\sim\,100$ (bulk) above the dipole-limit floor, with the surface and bulk $\chi^{(2)}$ contributions separately recoverable and the far from resonance response diamagnetic-dominated as the cold-plasma picture predicts.
At the oxygen K-edge of CO and atomic O on Ru(0001) on a conducting d-band surface, two observables separate the physics: the driven current reproduces the experimental near-edge lineshape~\cite{ostrom2015}, while the absorbed energy carries a purely absorptive beyond-dipole correction of $\sim 14\%$ which is the signature of the quadrupole-allowed $\mathrm{O}\,1s \to d$ channel that is non-trivial at the transition state geometry and for each isolated adsorbate.
This shows that all-electron full minimal coupling can follow the core-level response of a working catalytic interface, and it opens up predicting operando soft X-ray absorption of reaction intermediates at metal surfaces, where the quadrupole-allowed channels carry real beyond-dipole weight.
This capability could be further improved by coupling the electronic propagation to nuclear motion through an Ehrenfest framework, and it would enable time-resolved and pump--probe simulations.

\begin{suppinfo}
The Supporting Information presents the raw G-resolved current spectra underlying the thiophene S K-edge channel-crossover benchmark (Figure~S1); the surface-passivation dependence of the diamond forward second-harmonic generation benchmark, the forbidden-channel signal for four $(001)$ terminations (hydrogen, helium, and mixed) under full minimal coupling and in the dipole limit (Figure~S2 and Table~S1); and the beyond-dipole correction at the transition state and for the isolated species in the CO/Ru(0001) benchmark (Figures~S3 and S4).
\end{suppinfo}

\begin{acknowledgement}
This work was supported by the U.S.\ Department of Energy, Office of Science, Basic Energy Sciences, Chemical Sciences, Geosciences, and Biosciences Division under Contract DE-SC0023397 and DE-SC0023355.
This research used resources of the National Energy Research Scientific Computing Center, a DOE Office of Science User Facility supported by the Office of Science of the US DOE, under contract no.\ DE-AC02-05CH11231, using NERSC award BES-ERCAP0031431 and BES-ERCAP0035240.
This work also used the Expanse supercomputer at the San Diego Supercomputing Center through allocation PHY210131 from the Advanced Cyberinfrastructure Coordination Ecosystem: Services \& Support (ACCESS) program, which is supported by NSF Grant No.\ 2138259, No.\ 2138286, No.\ 2138307, No.\ 2137603, and No.\ 2138296.
\end{acknowledgement}

\section*{Data Availability Statement}
The FHI-aims input and output files for all benchmark calculations, together with the analysis scripts used to generate the figures, are available on zenodo.org at \href{http://dx.doi.org/10.5281/zenodo.22756160}{doi: \textcolor{blue}{10.5281/zenodo.22756160}}.
The full-minimal-coupling RT-TDDFT implementation described here is being contributed to the FHI-aims code base and will be available in a future release; until then it is available from the authors.

\bibliography{refs}
\end{document}